\documentclass[fleqn,10pt]{wlscirep}
\usepackage[utf8]{inputenc}
\usepackage[T1]{fontenc}
\usepackage{cleveref}
\usepackage{graphicx}
\usepackage{subcaption}
\usepackage{caption}
\usepackage{amsthm}
\newtheorem{definition}{Definition}
\usepackage{amsmath}

\title{Breakthrough Asymmetries across Disciplines and Countries: A Network approach to Structural Complexity of Scientific Progress}
\author[a,b,+]{Adarsh Raghuvanshi}
\author[c,+]{Hrishidev Unni}
\author[a,b]{Vinayak}
\author[c,*]{Anirban Chakraborti}
\affil[a]{CSIR-National Institute of Science Communication and Policy Research,
Dr. K.S. Krishnan Marg, New Delhi-110012, India}
\affil[b]{Academy of Scientific and Innovative Research (AcSIR), Ghaziabad-201002, India}
\affil[c]{School of Computational \& Integrative Sciences, Jawaharlal Nehru University, 
New Delhi-110067, India}

\affil[*]{anirban@jnu.ac.in}

\affil[+]{these authors contributed equally to this work}

\begin{abstract}

Science is driven by community endeavors across diverse fields and specializations, forming a complex structure that renders conventional performance evaluation methods inadequate. Using established indicators, the network-based normalized citation score, and the disruptive index, combined with the GENEPY algorithm, we evaluate the complexity rank of countries based on their breakthrough performance across 89 subfields of physical sciences, drawing on nearly 60 million articles (1900–2023). This quality-focused integrated approach reveals pronounced asymmetries: while countries such as the United States, Israel, and several in Europe sustain long-term structural advantages, emerging nations show rapid gains in later decades. A power-law relationship between aggregated breakthrough performance and countries' R\&D expenditure underscores the unequal and scale-dependent nature of global science. These results demonstrate that scientific advancement arises not from uniform growth but from asymmetric complexity, offering actionable insights for policymakers and funding agencies aiming to foster sustainable, high-quality research ecosystems.

\end{abstract}

\begin{document}

\flushbottom
\maketitle
%
%
\thispagestyle{empty}
\section*{Introduction}
Measurement of scientific performance is of central concern for agencies steering scientific enterprises with an overarching objective to strengthen scientific capacity, promote innovation, and align research outputs with broader socioeconomic and developmental priorities \cite{van2004measuring}. Since Francis Bacon's conception of science as a collective social endeavor, knowledge creation has been recognized as inherently social and embedded within broader societal structures \cite{jurgen2012francis}. Contemporary science is vast and evolving in complex structures of diverse fields and specializations, many of which converge to enable interdisciplinary advances \cite{dworking2019emergent}. It now functions as a complex, collaborative structure shaped by disciplinary interdependence and broader societal factors such as politics, economics, and culture. Countries and large research organizations typically demonstrate excellence not in a single domain but across multiple fields, reflecting both specialization and diversification. This makes performance evaluation nonlinear: diversity fosters competition and catalyzes knowledge growth through intertwined disciplinary networks, while traditional volume-based measures overlook the structural intricacies of scientific production, failing to account for quality, disciplinary norms, academic seniority, authorship practices, and other contextual factors \cite{kaur2015quality,fire2019over}, network-based approaches allow for more equitable, nuanced, and policy-relevant comparisons of scientific performance across heterogeneous research landscapes~\cite{Chakrabarti_DataScienceComplexSystems_2023}.

Publications and citations remain the dominant indicators of scientific productivity and impact \cite{braun1995scientometric, moed2005citation}. While publication counts capture research output, they emphasize volume over quality, and citation-based metrics, though reflective of recognition and knowledge diffusion across disciplines \cite{radicchi2011citation, chen2010community, zhao2015analysis}, are influenced by field-specific practices, publication age, and other biases \cite{van2004measuring, bornmann2008citation}.
Normalization is therefore essential for fair cross-field comparison \cite{waltman2015field, radicchi2012reverse, bornmann2020should}. More importantly, raw citation counts emphasize scale over substance, overlooking the role of high-quality work that drives scientific advancement. Breakthrough research, whether reinforcing existing knowledge driving cumulative advancement or disrupting them to open new frontiers, has long been understood as the actuator of scientific [r]evolution \cite{kuhn1997structure, kuhn1970nature, park2023papers, krauss2024debunking}. Identifying and distinguishing such contributions is thus critical for evaluating true scientific capacity. 

In this study, we analyze the domain of \emph{Physical Sciences}, encompassing 10 fields and 89 subfields over multiple decades of publication and citation data, providing the necessary resolution to track breakthrough performance and its evolution across heterogeneous research landscapes. We propose an integrated framework that combines two complementary indicators: the network-based normalized citation (NBNC) scores \cite{ke2023network} to identify the scientific breakthroughs and the disruptive index (CD) \cite{funk2017dynamic, wu2019large, park2023papers} to categorize these breakthroughs into consolidating and disruptive types. Both measures are derived from the local citation networks of publications, thereby embedding each contribution within its immediate knowledge context. 

Comparing the performance of countries across subfields still remains challenging. Composite indices that aggregate diverse metrics often depend on ad hoc weighting \cite{ioannidis2016multiple, forthmann2024summing}, failing to capture the structural complexities through which new capabilities emerge. An analogous problem is recognized in economics: GDP alone cannot capture a country's productive capacity because the composition of goods matters. This inspired complexity indices that link products to producers, providing a capability-based measure of development \cite{hidalgo2009building}. By analogy, we substitute subfields for products and adopt an interpretation that \emph{more complex countries contribute to more complex subfields, and more complex subfields are sustained only by more complex countries}. We formalize this relationship as a bipartite network of countries and subfields, with centrality measures computed through the Generalized Economic Complexity (GENEPY) scheme \cite{hidalgo2009building, tacchella2012new, sciarra2020reconciling}. This yields rankings that integrate diversity and exclusivity, and generate valuable insights needed to reshape policies that enhance the scientific capabilities and competence of macro enterprises.

\begin{figure}[htbp]
\centering
\includegraphics[width=0.98\linewidth]{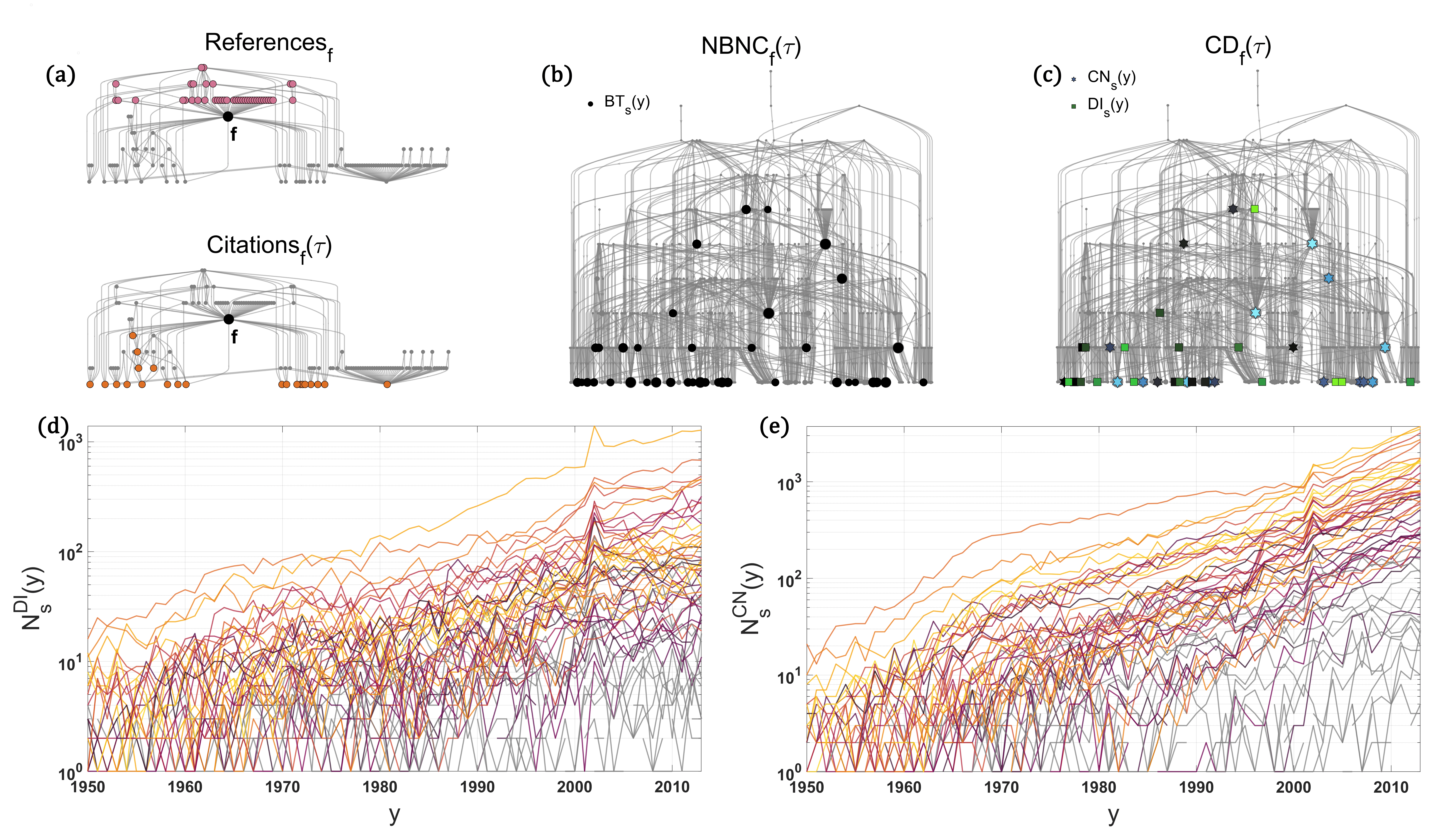}
\caption{
\textbf{Identification of consolidating and disruptive breakthroughs.} (\textbf{a}) An example node $f$ and its neighbors in a proxy citation network ($f$ highlighted in black) with $\mathrm{References}_f$ and $\mathrm{Citations}_f$ highlighted in the top and bottom rows respectively. (\textbf{b}) The larger proxy citation network representing the papers of a publication year $y$ for a subfield $s$ with node size proportional to $\mathrm{NBNC}_f(\tau)$; the top 5\% are selected as breakthroughs of that year ($\mathrm{BT}_s(y)$) and highlighted. Note that this network is a subgraph of the complete citation graph across all subfields and all years. Also, panels (\textbf{a}--\textbf{c}) are schematic illustrations serving as proxies: the networks shown are not drawn from the dataset, and the NBNC $\&$ CD values are randomly assigned to illustrate the procedure. (\textbf{c}) The same subgraph with $\mathrm{BT}_s(y)$ nodes color-coded by $\mathrm{CD}_f(\tau)$ (where $\mathrm{CD}_f(\tau) \leq 0$ towards blue and $\mathrm{CD}_f(\tau) > 0$ towards green) to indicate consolidating ($\mathrm{CD}_s(y)$) and disruptive ($\mathrm{DI}_s(y)$) types. (\textbf{d}) Yearly counts of disruptive breakthroughs ($N_s^{\mathrm{DI}}(y) = |\mathrm{DI}_s(y)|$) and (\textbf{e}) consolidating breakthroughs $N_s^{\mathrm{CN}}(y) = |\mathrm{CN}_s(y)|$ across 89 subfields.}
\label{fig:timeseries_BT_DI}
\end{figure}

\section*{Results}
\begin{figure*}[htbp]
\centering
\includegraphics[width=.85\linewidth]{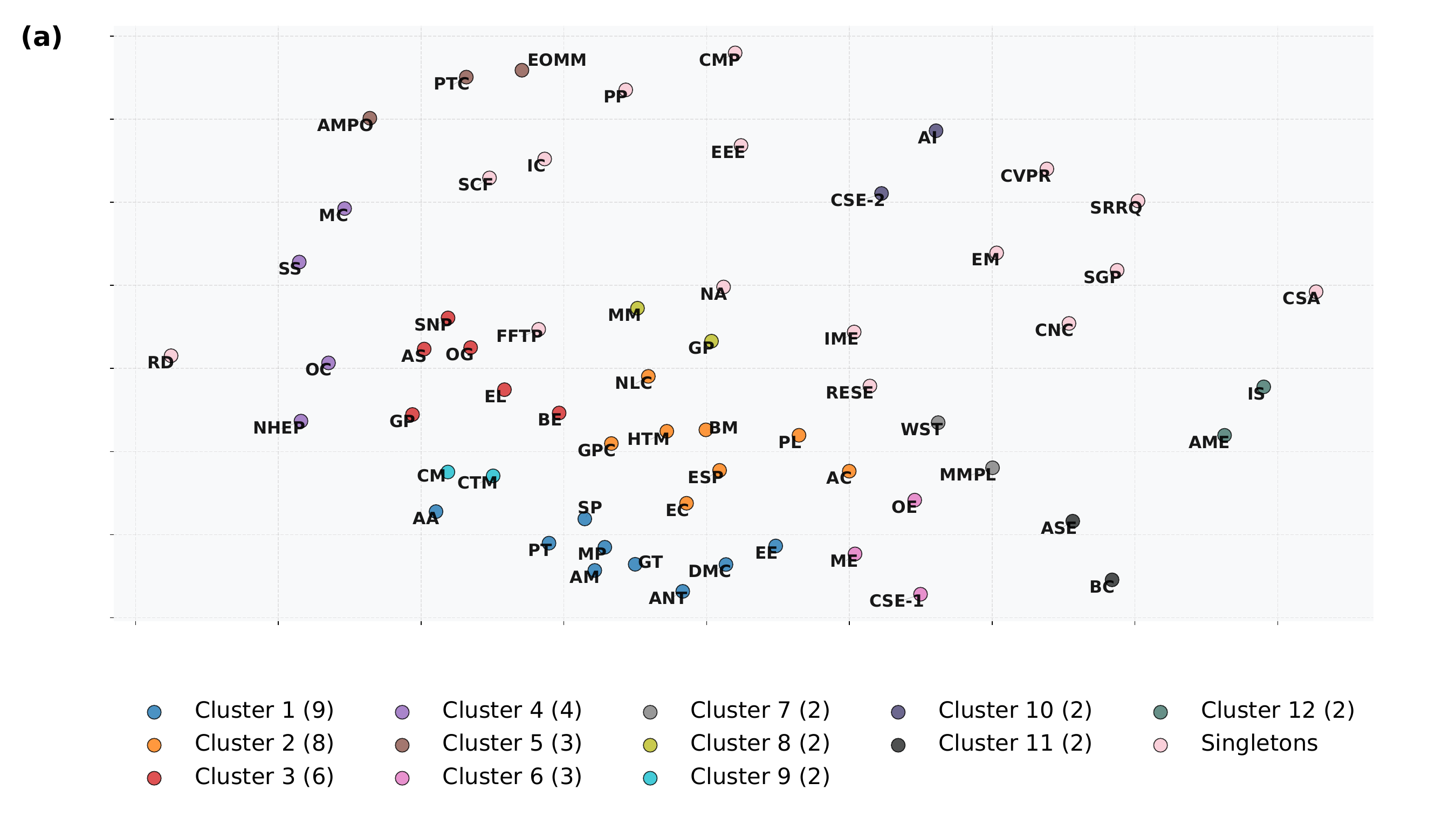}
\includegraphics[width=0.95\linewidth]{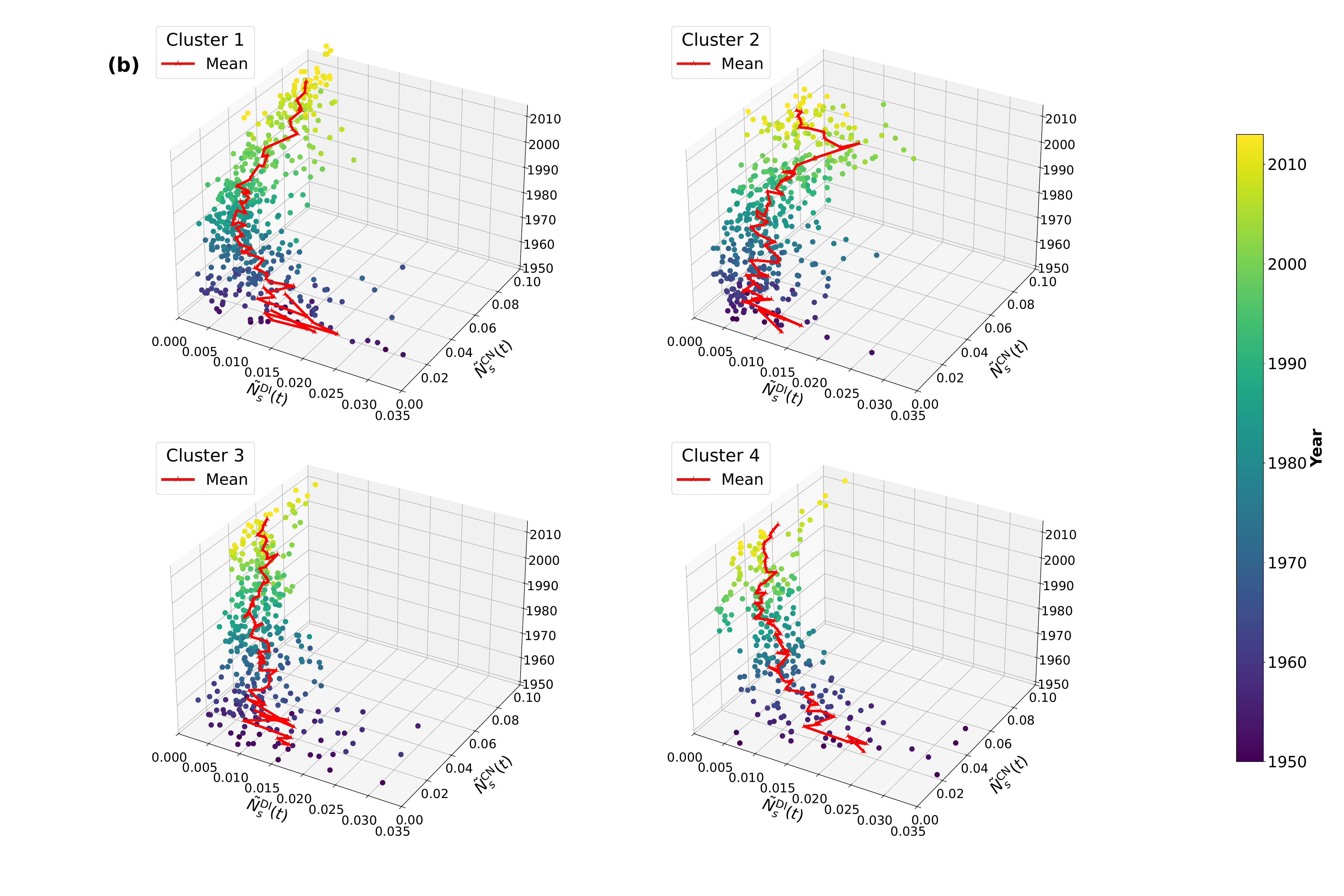}
\caption{
Clustering of subfields based on growth trajectories in scaled performance variables. {(\bf a)} Groups of subfields with similar growth trajectories in the scaled performance variables, $\tilde{N}^{\mathrm{CN}}_{s}(t)$ {\it vs }$\tilde{N}^{\mathrm{DI}}_{s}(t)$ (see main text), identified from 61 selected subfields ({\it SI}, Table \ref{tab:selected_subfields}). A dynamic time warping (DTW) distance was used to construct the similarity matrix, and the Leiden algorithm was applied to detect 12 clusters with size $\ge 2$ ({\it SI}, Table \ref{tab:clustered_fields}). {(\bf b)} Average growth trajectories of the four largest clusters, shown alongside their mean cluster trends. Trajectories of smaller clusters are presented in {\it SI}, Fig. \ref{fig:ClusterDynamics}.}
\label{fig:clusters}
\end{figure*}

We begin with approximately 60 million research articles in the Physical Sciences, drawn from the OpenAlex database and spanning the years 1900 to 2023. Annually, the top 5\% of publications from 1950–2013, as determined by the network-based normalization score (NBNC) that reduces the field and temporal citation biases, are identified as breakthrough papers (see Materials and Methods). 
Each breakthrough is further categorized as either consolidating (CN) or disruptive (DI), based on the CD index (CD; see Materials and Methods). This classification yields time series of consolidating and disruptive breakthrough-counts for each of the 89 subfields represented in the dataset.

As the total number of publications grows exponentially over time \cite{price1963little, bornmann2021growth}, the absolute count of breakthroughs also increases, although their relative shares remain informative. When normalized by annual publication volume in each subfield, the resulting trajectories capture the proportion of papers that qualify as consolidating or disruptive breakthroughs, as shown respectively in Figs.~\ref{fig:timeseries_BT_DI}d and ~\ref{fig:timeseries_BT_DI}e. These trajectories reveal substantial heterogeneity across subfields, with weak correlations between CN and DI trends within the same subfield. To compare the temporal dynamics of subfields, we clustered these normalized breakthrough trajectories using dynamic time warping (DTW) similarity \cite{keogh2005exact} and the Leiden algorithm \cite{traag2019louvain}(see Materials and Methods). From the 89 subfields, we retained 61 with complete and consistent time series profiles of publication volume and breakthrough proportions, which partition into 12 multi-subfield clusters and 16 singletons (Table~\ref{tab:clustered_fields}). Four large clusters account for nearly half of the retained subfields (Fig.~\ref{fig:clusters}a).  Representative average trajectories, along with the individual growth trajectories of member subfields, are shown in Fig.~\ref{fig:clusters}b for the 4 largest clusters, illustrating distinct temporal patterns across groups. Cluster 1, which includes {Astronomy}, {Paleontology}, Environmental Engineering, and 6 other from the subfields of \emph{Mathematics}, exhibits an early dominance of disruption peaking in the 1960s, followed by consolidation through the 1970s and 1980s and a modest resurgence of disruption after 2000. Cluster 2, encompassing {Analytical Chemistry}, {Global and Planetary Change}, and {Environmental Chemistry}, along with several other in {Environmental Science} and health-related subfields, shows a similar pattern but with a stronger resurgence of disruption in the 1990s. In contrast, Clusters 3 and 4 transition earlier to consolidation and remain relatively stable. These results demonstrate that subfields follow distinct evolutionary paths, with some subfields cycling between disruption and consolidation while others stabilize more quickly (See Fig.~\ref{fig:ClusterDynamics} for the other clusters).

\begin{figure}[htbp]
\centering
\includegraphics[width=0.9\linewidth]{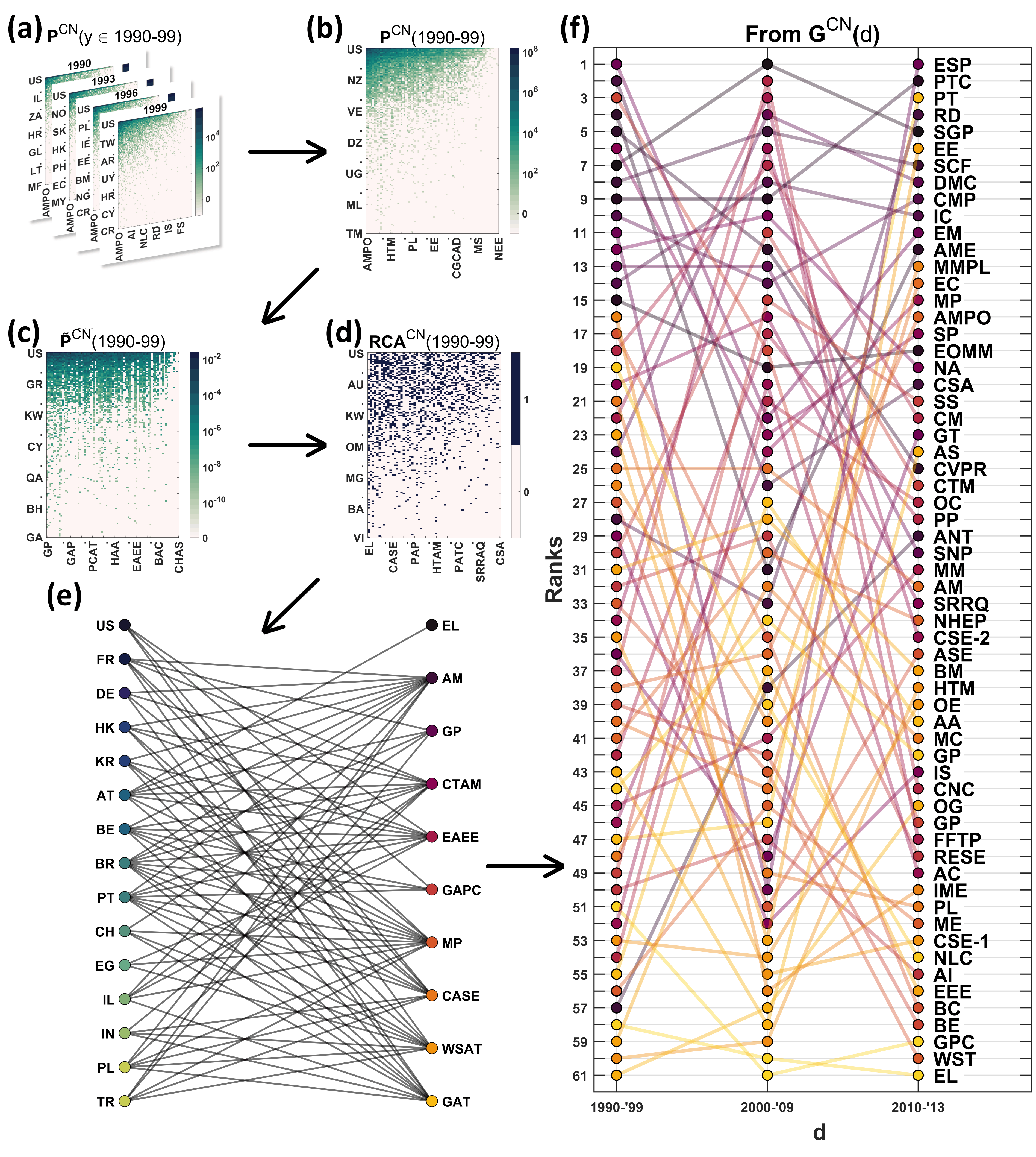}
\caption{\textbf{Construction of the country-subfield bipartite network.} (\textbf{a}) For each year $y$, consolidating breakthroughs are attributed to countries and subfields to construct matrices $\textbf{P}^{\mathrm{CN}}(y)$, where each element gives the number of breakthroughs in a given country-subfield pair. Heatmaps of these matrices for four years ($y \in \{ 1990,1993,1996,1999\}$) in the decade $1990-99$ are shown. (\textbf{b}) Yearly matrices are aggregated into decade-wise counts to get $\textbf{P}^{\mathrm{CN}}(1990-99)$ and visualised similarly. (\textbf{c}) The decade-wise matrix is normalised by the total number of breakthroughs in each subfield to obtain $\tilde{\textbf{P}}^{\mathrm{CN}}(1990-99)$ and visualised. (\textbf{d}) After calculating the $\textbf{RCA}^{\mathrm{CN}}(d)$ a threshold is applied to obtain the adjacency matrix $\textbf{A}^{\mathrm{CN}}(1990-99)$. (\textbf{e}) Visualisation of the resulting country-subfield bipartite network represented by this adjacency matrix for a subset of nodes. (\textbf{f}) Competitive rankings for 61 subfields over three consecutive decades, namely the $5^{th}$, $6^{th}$, and $7^{th}$ decades, are presented. The rankings are shown on the left ordinate, while the subfields are listed on the right ordinate. A comprehensive overview of all decades, highlighting both consolidating and disruptive breakthroughs, is displayed in {\it SI}, Fig. \ref{fig:Decadal_Subfield_Ranking}.}
\label{fig:bipartite}
\end{figure}

We next investigate how breakthroughs are distributed across countries and subfields. Using author affiliations, we construct decade-level country-subfield matrices of consolidating and disruptive breakthroughs. For a comparative assessment of the significance of the contribution of a country in a subfield, we use a revealed comparative advantage (RCA) filter introduced in econometrics \cite{balassa1965trade}, measuring a country’s relative performance in producing breakthroughs within a specific subfield, compared to that subfields' overall contribution to global breakthrough production. A threshold criterion helps distinguish countries with and without comparative advantage; those exceeding the threshold in a subfield are considered competitive performers, while those below it are not. The RCA filtering results in binary adjacency matrices that indicate whether a country is structurally specialized in a subfield during a given decade. Early decades show RCA concentrated in a few countries, but over time, broader participation emerges (Figs.~\ref{fig:bipartite}d,~\ref{fig:MatrixplotCN},~\ref{fig:MatrixplotDI}). By the 2000s, many subfields, including {Geophysics}, {Atmospheric Science}, and {Global and Planetary Change}, exhibit near-universal RCA coverage, while others, such as {Water Science and Technology}, witnessed broader adoption in later decades. This triangular structure, in which less diversified countries specialize only in ubiquitous subfields, mirrors patterns in economic complexity analysis \cite{tacchella2012new, cimini2014scientific}.

\begin{figure}[htbp]
  \centering
    \includegraphics[width=.98\linewidth]{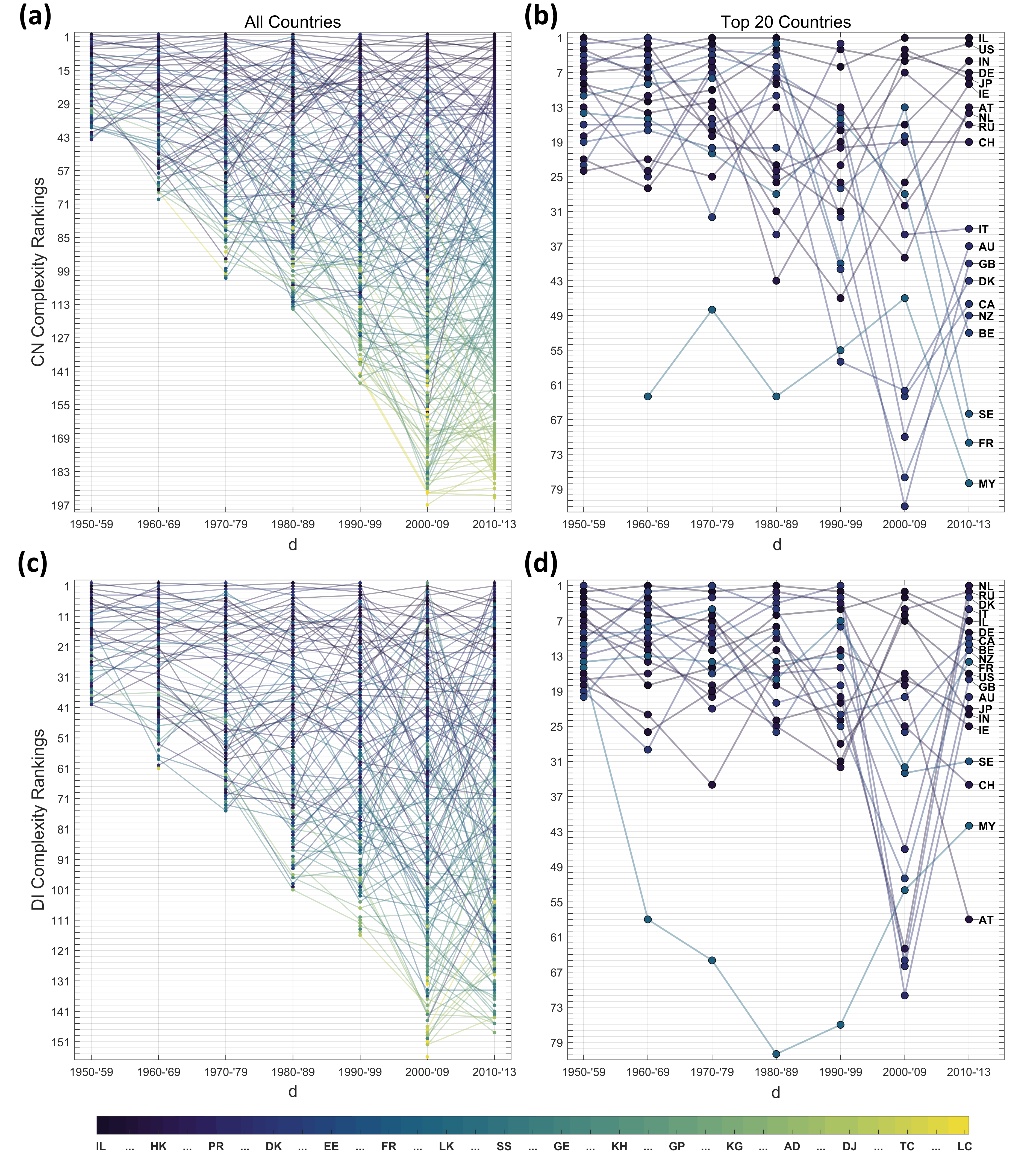}

  \caption{\textbf{The decadal evolution of complexity rankings.} The ranks calculated for the countries using the two vectors $\textbf{H}^{CN}(d)$ and $\textbf{H}^{CN}(d)$ (see {\bf Materials and Methods}), covering the period from $1950$ to $2013$ are plotted. The evolution of the ranks of all the countries is shown in {\bf (a)} for disruptive breakthroughs, and in {\bf (c)} for consolidating breakthroughs. The rank trajectories of the countries ranked in the top 20 of consolidating breakthroughs in the $1^{st}$ decade are highlighted in {\bf (b)} and {\bf (d)} for consolidating and disruptive breakthroughs, respectively.
  The countries are denoted by the country codes using the {\it alpha}-2 system and a colourmap. This is shown in the colorbar at the bottom.}
  \label{fig:Allcountry_ranks}
\end{figure}

Applying the GENEPY algorithm \cite{sciarra2020reconciling} to the RCA-filtered matrices yields complexity scores for both subfields and countries, facilitating a ranking of subfields by their structural complexity, the higher the structural complexity, the higher the rank, and of countries performing in the most complex subfields, in alignment with our circular interpretation. Subfield complexity rankings highlight two key trends. First, certain subfields, including {Signal Processing}, {Automotive Engineering}, and {Discrete Mathematics and Combinatorics}, retain high complexity rankings across decades in both DI and CN, reflecting concentration in scientifically diversified countries. Second, other subfields undergo marked shifts: Information Systems declines from a top-ranked position in the 1950s to widespread adoption by the 2000s, while {Condensed Matter Physics} and {Surface, Coating and Films} rise from peripheral positions into the top bracket. Combining subfield complexity rankings with their temporal dynamics reveals that {Automotive Engineering} maintained a high complexity ranking, unlike {Information Systems}; both belong to cluster 11 (Fig.~\ref{fig:ClusterDynamics}), reflecting concentrated contributions versus broad adoption. {Signal Processing} and {Discrete Mathematics and Combinatorics} also rank among the top in CN complexity, with {Signal Processing} identified as a singleton. Similarly, {Condensed Matter Physics} and {Surfaces, Coatings, and Films} move into the top bracket in later decades but remain singletons. In contrast, {Computer Networks and Communications} is also identified as a singleton but declined in the DI complexity ranking. These trajectories illustrate that GENEPY complexity reflects structural exclusivity: subfields concentrated among scientifically diversified countries maintain high scores, while those that diffuse broadly lose them. Importantly, consolidating and disruptive rankings are not always aligned: {Ecological Modelling} and {Geochemistry} rank highly for disruption but not consolidation, underscoring that structural exclusivity differs by breakthrough type. 

\begin{figure}[htbp]
\centering
\includegraphics[width=0.9\linewidth]{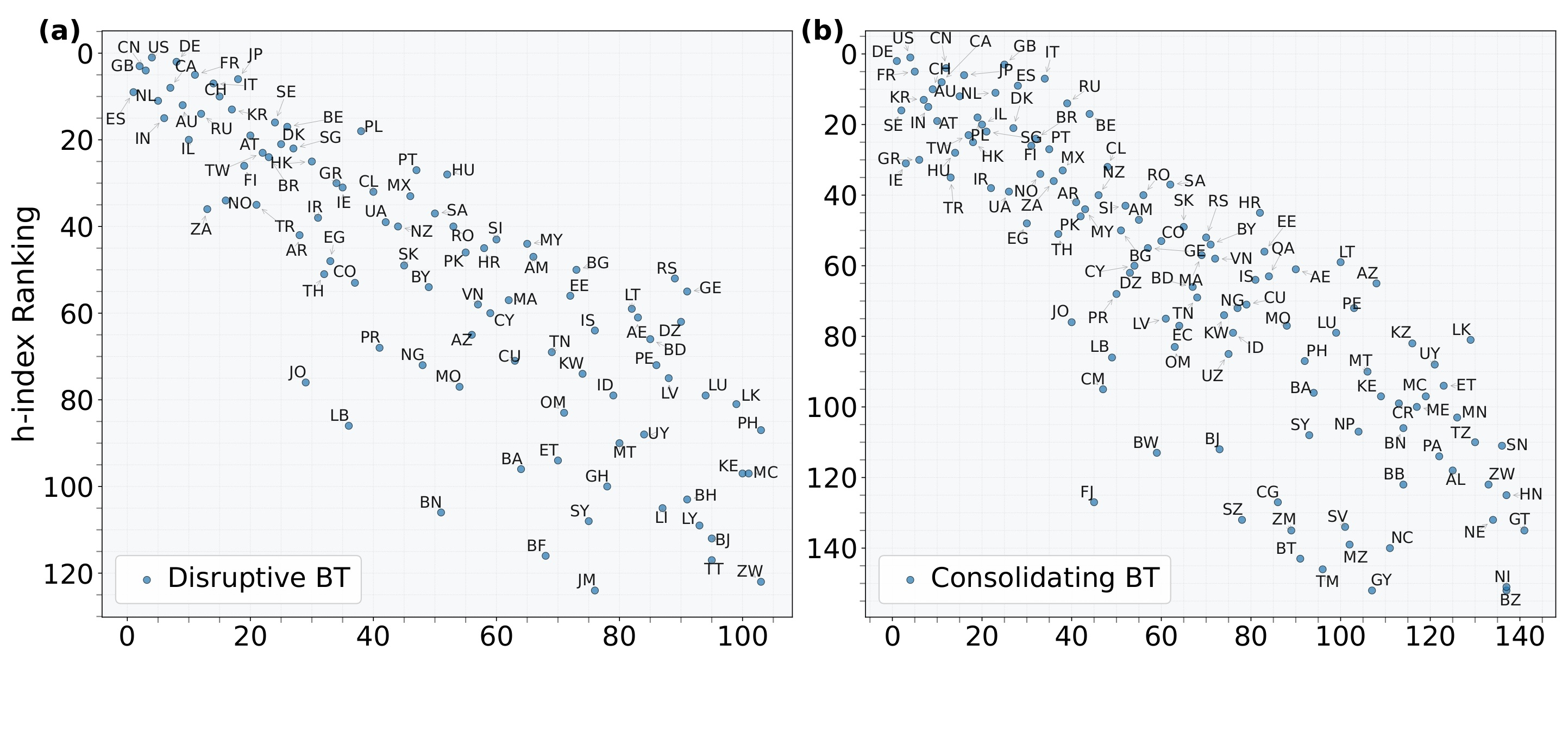}\llap{\parbox[b]{2in}{\textbf{Complexity Ranking}\\\rule{0ex}{-4in}}}
\llap{\parbox[b]{5.1in}{\textbf{Complexity Ranking}\\\rule{0ex}{-4in}}}
\caption{The correlation scatter plot between ranking of the countries based on their complexity scores: {\bf (a)} for disruptive breakthroughs, and {\bf (b)} for the consolidating breakthroughs, and ranking based on the $h$-index (Scimago) in the year 2005.}
\label{fig:Comp_Scimago}
\end{figure}

\begin{figure}[htbp]
\centering
\includegraphics[width=0.95\linewidth]{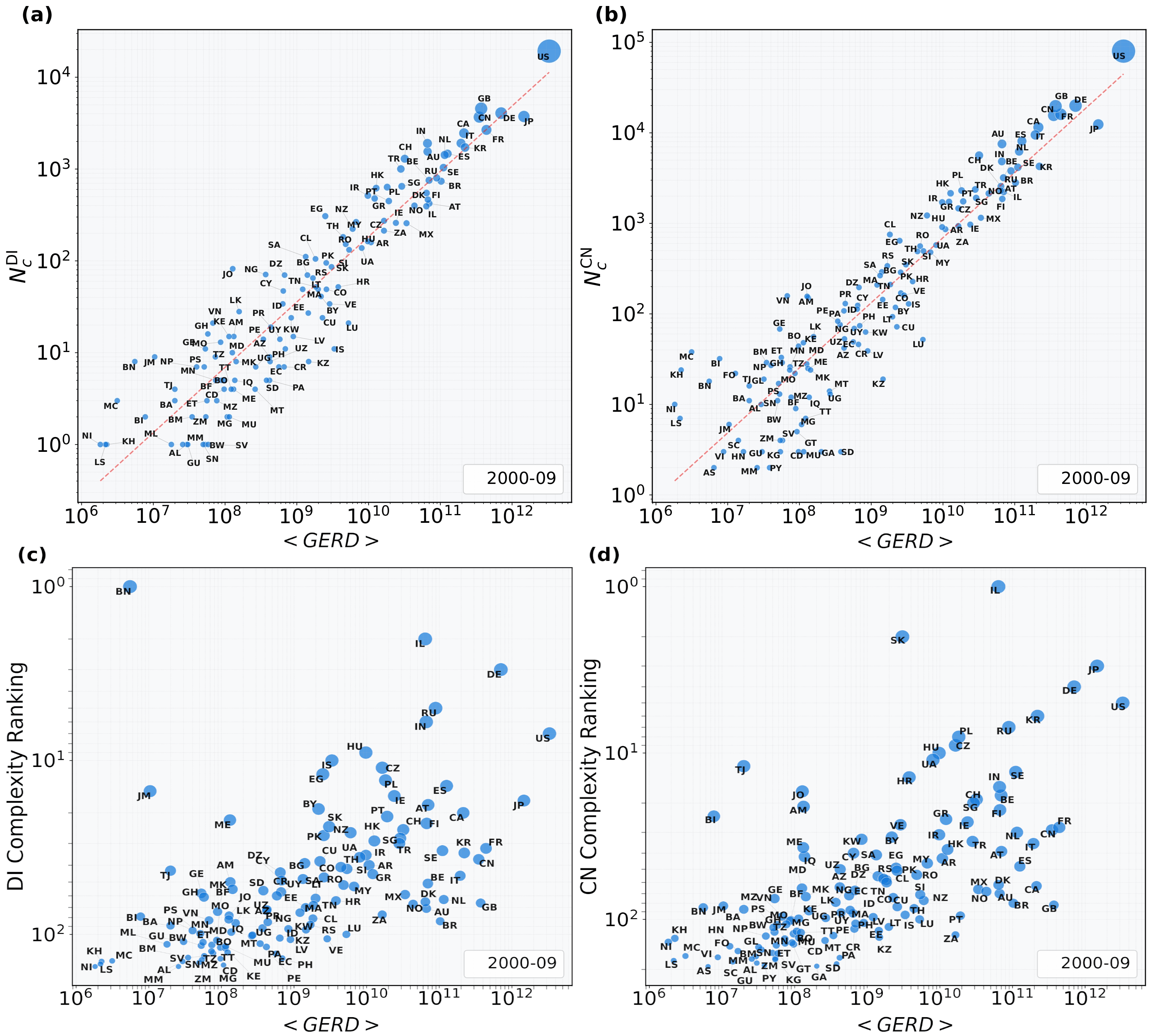}
\caption{The scatter plot of the GERD and the performances of countries in terms of total number of breakthroughs (top panel) and complexity rankings (bottom panel), over a decade. For $d=2000-09$ {\bf (a)} illustrates $N^{\mathrm{CN}}_{c}(d)\sim \langle GERD \rangle_{c}(d)$, and {\bf (b)} shows $N^{\mathrm{DI}}_{c}(d)\sim \langle GERD \rangle_{c}(d)$, where $N^{\mathrm{CN}}_{c}(d)$ and $N^{\mathrm{DI}}_{c}(d)$, are respectively the total number of consolidating and disruptive breakthroughs, and $\langle GERD \rangle_{c}(d)$ is the cumulative GERD of the country $c$ in the decade $d = 2000-09$. The relationship between $\langle GERD \rangle_{c}(d)$ and the competitive ranks of the countries is shown, respectively, in {\bf (c)} for disruptive breakthroughs, and {\bf (d)} for consolidating breakthroughs.}
\label{fig:GERD_relation_Perform}
\end{figure}

At the country level, complexity rankings capture long-term trajectories and volatility (Fig.~\ref{fig:Allcountry_ranks}). The United States remains consistently in the top bracket for both consolidating and disruptive breakthroughs. European countries show more varied patterns: Germany maintains stability, while France and Great Britain decline in consolidation after the 2000s. Israel exhibits sustained ascent, leading to consolidation by the 2010s. East and Southeast Asia, absent from early rankings, rise sharply from the 1990s onward, with China, Korea, and Singapore entering the global top ten. In contrast, several South American and African countries decline over time, though exceptions exist, notably Algeria, in consolidation. Volatility is also evident in Canada, Italy, and Vietnam, which show mid-period rises followed by declines. These trajectories reveal that while some countries maintain long-term structural advantage, others experience significant fluctuations, reflecting the interplay of institutional capacity and emerging infrastructures. Notably, despite volatility sustained with changes in threshold values, the United States, Israel, and some European countries consistently dominate the upper ranks (see Materials and Methods). We compare our complexity-based country rankings and conventional impact measures by comparing them with Scimago \( \mathsf{h} \)-index(\ref{hindex}) rankings for \emph{Physics and Astronomy} in 2005 (Fig.~\ref{fig:Comp_Scimago}). Both consolidating and disruptive breakthrough rankings exhibit modest positive correlations with \( \mathsf{h} \)-index rankings, with Spearman coefficients fluctuating between 0.1 and 0.4 during the period 1996--2013 (Fig.~\ref{fig:ScimagoRanking}).

Finally, we assess the relationship between breakthrough performance and research investment. Total breakthrough counts scale, i.e., the linear approach, as a power-law with countries gross expenditure on R\&D (GERD), consistent with investment supporting aggregate output (Fig.~\ref{fig:GERD_relation_Perform}a and~\ref{fig:GERD_relation_Perform}b). However, complexity-based rankings exhibit a highly nonlinear relationship with GERD  (Fig.~\ref{fig:GERD_relation_Perform}c and~\ref{fig:GERD_relation_Perform}d). High expenditure does not guarantee high structural positioning; however, countries with higher per capita R\&D expenditure, such as Israel and Slovakia, are found as outliers, while small economies, viz., Brunei and Jamaica achieved high rankings in disruptive breakthroughs by collaborating with high-performing countries in complex subfields. In contrast, several European countries with substantial GERD fail to sustain top positions in consolidating breakthroughs over time. This pattern reflects the GENEPY ranking, which penalizes diffuse portfolios and promotes exclusivity. This divergence indicates that investment alone is insufficient to confer structural advantage; instead, specialization trajectories and systemic positioning in the global scientific network determine whether spending translates into long-term complexity gains. 
Overall, these results show that scientific subfields evolve along heterogeneous breakthrough trajectories, and that countries' positions in this landscape depend less on raw output or spending than on how effectively they specialize in structurally complex areas. This framework provides a prospective lens on the geography of science, highlighting where countries consolidate established strengths and where they develop footholds in frontier subfields.

\section*{Discussion}

Citations are widely used to gauge research impact, their multidimensional character complicates interpretations \cite{moed2005citation,leydesdorff2016citations}. Meaningful like-with-like comparison in ranking knowledge producers rests upon eliminating citation biases \cite{bornmann2019normalisation,vaccario2024quest}, we advance this debate by integrating the NBNC approach, the CD index, and the GENEPY method to rank countries across diverse subfields of the physical sciences. Unlike text-based approaches \cite{shibayama2021measuring,leahey2023types}, this framework builds on citation dynamics and, a priori, does not assume that all subfields have equivalent breakthrough potential, thereby serving the aims of this study. By doing so, our framework offers a coherent lens for measuring the asymmetric evolution of subfields. Our approach contributes two methodological advances extending the existing approaches. First, integration of the NBNC metric and CD index facilitates identification of scientific breakthroughs in consolidating and disruptive categories while maintaining sensitivity to subfield-specific dynamics. Second, using GENEPY-based ranking, we move beyond volume or impact counts toward structural exclusivity: the concentration of subfield breakthroughs across countries with diversified scientific portfolios. This ranking enables the tracing of countries' performance while incorporating specialization and competitive asymmetry, nuances that linear counting fails to capture. Importantly, the method is generalizable to meso-level system, such as research organizations and universities, providing a scalable tool to analyze structural complexity across multiple layers of scientific production. 

Our analysis rests on bibliometric data from OpenAlex, which provides broad coverage across decades \cite{culbert2025reference}. Nevertheless, important caveats remain. Missing affiliations in bibliometric data from periods predating the 1990s, the full counting approach for mapping counties, the inclusion of self-citations in computation, and sensitivity to threshold are all potential sources of bias. Subfields grouped in the same clusters appear to have co-evolved; their paradigmatic overlapping is beyond the scope of this work. Similarly, the causal relationship between the evolution of sub-fields is not addressed here. Volatility in subfield rankings, particularly for highly disruptive but sparsely populated breakthroughs, may reflect data sparsity or classification shifts as much as genuine structural change. These limitations do not undermine the robustness of long-run patterns but do warrant caution in interpreting short-term fluctuations or fine-grained rank changes. Despite these limitations, our analysis yields several insightful findings. The USA, Israel, and Germany maintain nearly a consistent profile in complexity rankings, underlining the sustainability of their research ecosystems in both consolidating and disruptive categories. European countries demonstrate comeback trajectories, exhibiting their resilient capacity in generating scientific breakthroughs. Breakthrough performance scales nonlinearly with countries’ GERD, emphasizing the role of investment in advancing knowledge, with the GERD–complexity relationship being highly nonlinear, with some outliers, which remain robust to threshold variations (Materials and Methods). It remains to understand how the increasing R\&D investment and funding strategies helped strengthen the research performance of developing countries, especially China and India. 

Our results reposition citation analysis: from measuring impact retrospectively to mapping the structural evolution of science prospectively. The framework introduced here, is not a substitute for existing indicators, but a complementary approach that highlights where countries build unique advantages in frontier subfields. For policymakers, this opens the possibility of targeting investment not simply to expand output, but to cultivate resilient ecosystems that sustain complex, high-barrier research trajectories. 
These findings highlight the complexity, underscoring the need for policies that cultivate specialization and resilient ecosystems rather than volume alone. Scientific competitiveness depends on sustaining complex, high-barrier research. Policies should: target specialization in frontier subfield(s) rather than expanding volume, build resilient ecosystems conducive to both consolidating and disruptive research \cite{pitak2012morality,bammer2008enhancing}, foster academic freedom, collaborative norms, and ethical research practices \cite{spannagel2023academic, pitak2012morality}, and optimize investment efficiency, leveraging the nonlinear GERD–complexity relationship.

\section*{Materials and Methods}
\label{sec:methods}
\subsection*{Data}
The data is extracted from the OpenAlex of approximately 60 million articles published from 1900 to 2023 in the Physical Sciences domain, encompassing 10 fields, viz. Engineering, Computer Science, Environmental Science, Physics $\&$ Astronomy, Material Science, Chemistry, Earth and Planetary Science, Mathematics, Energy, and Chemical Engineering. The $NBNC$ score and $CD$ index were computed from the corpus of all DOI publications. The NBNCs were computed from 1950 to 2013 for nearly 36 million articles over nearly 711 million citations.  

\subsection*{Breakthroughs Identification}
From the perspective of scientific innovation theory, breakthroughs can broadly be categorized into two types: consolidating (which reinforce and extend established paradigms), and disruptive (which challenge existing trajectories and open new research frontiers). To capture these dynamics empirically, we combine the network-based normalization citation (NBNC) scores, using which we identify the scientific breakthroughs, and the disruptive index (CD) to categorize these breakthroughs into consolidating and disruptive types. Both measures are derived from the local citation networks of publications, thereby embedding each contribution within its immediate knowledge context. The NBNC approach reduces the field bias while differing in the underlying assumption from the other normalization methods \cite{waltman2016review, chatterjee2016universality, radicchi2008universality, uzzi2013atypical}, as it does not implicitly assume uniform performance rates across the various subfields \cite{ke2023network}. Breakthrough (BT) papers are identified by selecting the top $k\%$ of publications ranked with respect to their NBNCs. For each publication year $y \in Y$, where Y={1900,\dots,2023}, we consider \( N_{s}^{\mathrm{BT}}(y) = \left|\{\,\text{papers in subfield } s \text{ and year } y \text{ with NBNC in top }k\% \,\}\right|.\) Each breakthrough is further categorized using the CD index, as consolidating (CN) for $\textrm{CD}<0$ and disruptive (DI) for $\textrm{CD}\ge 0$. Formally, \(N_s^{\mathrm{BT}}(y) = N_s^{\mathrm{CN}}(y) + N_s^{\mathrm{DI}}(y).\) In other words, we do not use CD as a metric for ranking; rather, we use it as a filter (see Supplementary for detailed explanation). 

Regarding parameters, we adopt a time horizon $\tau=10$ years, consistent with prior work \cite{ke2023network, uzzi2013atypical, redner2005citation} for computing NBNCs. Although this period is sufficiently long for most subfields, it may overlook notable cases in which publications remain uncited for extended periods before gaining recognition (“sleeping beauties”). For computing CD, we stay consistent with $\tau$. With the setting chosen in the prior study \cite{ke2023network}, we consider the threshold $k=5$.  

This framework helps identify not only notable contributions but also their impact type. For instance, twin articles on Renormalization Group and Critical Phenomena \cite{wilson1971renormalizationa, wilson1971renormalizationb}, which led K. G. Wilson to win the 1982 Physics Nobel prize, are identified as breakthroughs but consolidating type. In contrast, the foundation work \cite{kadanoff1966scaling} by Kadanoff is recognized as a disruptive breakthrough. Similarly, the work on Giant Magnetoresistance \cite{baibich1988giant} is identified as a disruptive breakthrough. On the other hand, gauging impactful works, typically categorized as sleeping beauties, like Higgs Bosons \cite{higgs1964broken}, requires a longer time window to be identified as a breakthrough.

\subsection*{Clustering of Breakthrough Trajectories}
As the publication output grows exponentially \cite{price1963little,bornmann2021growth}, the breakthrough counts also increase. As the publication patterns differ across subfields, we normalize the absolute breakthrough counts by annual subfield volume, as
$
\tilde{N}^{\mathrm{CN}}_s(y) = \frac{N^{\mathrm{CN}}_s(y)}{N_s(y)}$ and 
$
\tilde{N}^{\mathrm{DI}}_s(y) = \frac{N^{\mathrm{DI}}_s(y)}{N_s(y)},
$
where ${N_s(y)}$ is the total number of articles in subfield $s$ and year $y$. To study the temporal patterns, we treat the sequence of ordered pairs ($\tilde{N}^{\mathrm{CN}}_s(y)$,$\tilde{N}^{\mathrm{DI}}_s(y)$) as a breakthrough trajectory. The distance function could have been chosen to be Euclidean or Correlation, but both fail to account for the temporal gaps between the subfield evolutions. To address this, we use the dynamic time warping (DTW), which aligns sequences, allowing nonlinear temporal shifts. Distances are transformed into similarities via a Gaussian kernel, producing a weighted subfield network. Clusters are identified using the Leiden community detection algorithm \cite{traag2019louvain}. Subfields with incomplete or unstable trajectories are excluded (61 of 89 retained).
\subsection*{Complexity Scores}
For each decade $d \in D = \{1950-59,1960-69,\dots,2010-13\}$, we construct the matrices
$\textbf{P}^{\alpha} (d)=[P_{cs} ^{\alpha} (d)]$, 
where $\alpha \in \{\mathrm{CN,DI}\}$, and $P_{cs}^{\alpha} (d)$ is the number of $\alpha$-type breakthroughs from country c in subfield s. The absolute counts are normalized by the total number of breakthroughs during that period to calculate $ \tilde{\textbf{P}}^{\alpha} (d)=[\tilde{P}_{cs} ^{\alpha} (d)]$, where $\tilde{P}_{cs} ^{\alpha} (d) = \frac{P_{cs} ^{\alpha} (d)}{\sum_{y \in d}N_s^{BT}(y)}$. After applying revealed comparative advantage (RCA) on $\tilde{\textbf{P}}^{\alpha}$ as,
$
\mathrm{RCA}^\alpha_{cs}(d) = \frac{\tilde{P}^\alpha_{cs}(d)}{\sum_{s'} \tilde{P}^\alpha_{cs'}(d)} \Big/ \frac{\sum_{c'} \tilde{P}^\alpha_{c's}(d)}{\sum_{c',s'} \tilde{P}^\alpha_{c's'}(d)},
$
we construct an adjacency matrix $\textbf{A}^\alpha(d) = [A^{\alpha}_{cs}(d)]$, by setting $A_{cs}^{\alpha}(d)=1$ if $RCA_{cs}^{\alpha}(d)\geq 1$, and 0 otherwise, yielding binary adjacency matrices $\textbf{A}^{\alpha}(d)$ representing bipartite networks between countries and subfields. (The RCA reference value $1$ represents the world average performance; only countries scoring RCA values equal to or above $1$ are chosen as performers.) Two sets of "complexity" scores for countries ($\textbf{H}^\alpha(d) = [H^\alpha_{c}(d)]$) and subfields ($\textbf{G}^\alpha(d) = [G^\alpha_{s}(d)]$) are derived by applying the GENEPY algorithm \cite{sciarra2020reconciling} to the $\textbf{A}^{\alpha}(d)$ (see Supplementary\ref{rca_genepy} for explanation).

\subsection*{Sensitivity of results to threshold parameter}

The complexity ranking depends on the threshold $k$, which defines the standard for breakthroughs. Results across $1\%$, $5\%$, and $10\%$ thresholds were highly consistent: country rankings from $N_{c}^{\alpha}(d)$ correlated strongly ($\rho \gtrsim 0.9$), and exponent values in Fig.~\ref{fig:GERD_relation_Perform} varied by only $\sim \pm 3\%$. For $N_{c}^{\mathrm{DI}}(d)/N_{c}^{\mathrm{CN}}(d)$, rank correlations were $\sim0.91$, $\gtrsim0.87$, and $\sim0.72$ for threshold pairs $\{5\%,10\%\}$, $\{1\%,5\%\}$, and $\{1\%,10\%\}$. Subfield ratios $N_{s}^{\mathrm{DI}}(d)/N_{s}^{\mathrm{CN}}(d)$ showed lower stability ($0.63$–$0.70$), indicating greater sensitivity at finer scales. 

Triangularity in $\mathsf{A}^{\alpha}(d)$ is most pronounced for $k=1$, outperforming $k=5,10$. Lower thresholds improved ranks for high-GERD countries in later decades, while China and India's rankings are marginally benefited from higher thresholds. The high-ranking small economies remained stable, though Brunei lost its lead at low thresholds. The sixth-decade dip attenuated or disappeared (e.g., Great Britain for DI ranking) under low thresholds. 

Overall, aggregate results are robust to $k$, though complexity indicators are more threshold-sensitive. All correlations are Spearmans' $\rho$.

\subsection*{Data Availability}
Bibliometric data of the articles and reviews from the domain of {\it Physical Sciences} have been extracted from the OpenAlex electronic database, from 1900 to 2023 \url{(https://openalex.org/)}. The Scimago ranking of the countries was obtained from the Scimago database \url{(https://www.scimagojr.com/countryrank.php)} for the field {\it Physics and Astronomy}. The R\&D expenditure (in percentage) and GDP data have been extracted from the "World Bank" database,  respectively from \url{(https://data.worldbank.org/indicator/GB.XPD.RSDV.GD.ZS)} and \url{(https://data.worldbank.org/indicator/NY.GDP.MKTP.CD)}, for computing GERD.

\subsection*{Acknowledments}
Adarsh Raghuvanshi acknowledges the University Grant Commission, India (NTA Ref. No.: 191620064987) for financial support.

\bibliography{referencesf}

@book{Chakrabarti_DataScienceComplexSystems_2023,
  author    = {Chakrabarti, Anindya S. and Bakar, K. Shuvo and Chakraborti, Anirban},
  title     = {Data Science for Complex Systems},
  publisher = {Cambridge University Press},
  address   = {Cambridge},
  year      = {2023},
  isbn      = {9781108953597},
  doi       = {10.1017/9781108953597},
  url       = {https://www.cambridge.org/core/books/data-science-for-complex-systems/304F66053C62CD439FDFA46D2D4323A8}
}

@article{fire2019over,
  title={Over-optimization of academic publishing metrics: observing Goodhart's Law in action},
  author={Fire, Michael and Guestrin, Carlos},
  journal={GigaScience},
  volume={8},
  number={6},
  pages={giz053},
  year={2019},
  publisher={Oxford University Press}
}

@article{kaur2015quality,
  title={Quality versus quantity in scientific impact},
  author={Kaur, Jasleen and Ferrara, Emilio and Menczer, Filippo and Flammini, Alessandro and Radicchi, Filippo},
  journal={Journal of Informetrics},
  volume={9},
  number={4},
  pages={800--808},
  year={2015},
  publisher={Elsevier}
}

@article{uzzi2013atypical,
  title={Atypical combinations and scientific impact},
  author={Uzzi, Brian and Mukherjee, Satyam and Stringer, Michael and Jones, Ben},
  journal={Science},
  volume={342},
  number={6157},
  pages={468--472},
  year={2013},
  publisher={American Association for the Advancement of Science}
}

@article{chatterjee2016universality,
  title={Universality of citation distributions for academic institutions and journals},
  author={Chatterjee, Arnab and Ghosh, Asim and Chakrabarti, Bikas K},
  journal={PloS one},
  volume={11},
  number={1},
  pages={e0146762},
  year={2016},
  publisher={Public Library of Science San Francisco, CA USA}
}

@article{waltman2016review,
  title={A review of the literature on citation impact indicators},
  author={Waltman, Ludo},
  journal={Journal of informetrics},
  volume={10},
  number={2},
  pages={365--391},
  year={2016},
  publisher={Elsevier}
}

@article{balassa1965trade,
  title={Trade liberalisation and revealed comparative advantage},
  author={Balassa, Bela},
  journal={The manchester school},
  volume={33},
  number={2},
  pages={99--123},
  year={1965},
  publisher={Wiley Online Library}
}

@article{traag2019louvain,
  title={From Louvain to Leiden: guaranteeing well-connected communities},
  author={Traag, Vincent A and Waltman, Ludo and Van Eck, Nees Jan},
  journal={Scientific reports},
  volume={9},
  number={1},
  pages={1--12},
  year={2019},
  publisher={Nature Publishing Group}
}

@article{tacchella2012new,
  title={A new metrics for countries' fitness and products' complexity},
  author={Tacchella, Andrea and Cristelli, Matthieu and Caldarelli, Guido and Gabrielli, Andrea and Pietronero, Luciano},
  journal={Scientific reports},
  volume={2},
  number={1},
  pages={723},
  year={2012},
  publisher={Nature Publishing Group UK London}
}

@inproceedings{krauss2024debunking,
  title={Debunking revolutionary paradigm shifts: evidence of cumulative scientific progress across science},
  author={Krauss, Alexander},
  booktitle={Proceedings A},
  volume={480,no. 230},
  pages={20240141},
  year={2024},
  organization={The Royal Society}
}

@article{spannagel2023academic,
  title={The academic freedom index and its indicators: Introduction to new global time-series V-Dem data},
  author={Spannagel, Janika and Kinzelbach, Katrin},
  journal={Quality \& Quantity},
  volume={57},
  pages={3969--3989},
  year={2023},
  publisher={Springer}
}

@article{waltman2015field,
  title={Field-normalized citation impact indicators and the choice of an appropriate counting method},
  author={Waltman, Ludo and van Eck, Nees Jan},
  journal={Journal of Informetrics},
  volume={9},
  number={4},
  pages={872--894},
  year={2015},
  publisher={Elsevier}
}

@article{cimini2014scientific,
  title={The scientific competitiveness of nations},
  author={Cimini, Giulio and Gabrielli, Andrea and Sylos Labini, Francesco},
  journal={PloS one},
  volume={9},
  number={12},
  pages={e113470},
  year={2014},
  publisher={Public Library of Science San Francisco, USA}
}

@article{keogh2005exact,
  title={Exact indexing of dynamic time warping},
  author={Keogh, Eamonn and Ratanamahatana, Chotirat Ann},
  journal={Knowledge and information systems},
  volume={7},
  number={3},
  pages={358--386},
  year={2005},
  publisher={Springer}
}

@article{vaccario2024quest,
  title={The quest for an unbiased scientific impact indicator remains open},
  author={Vaccario, Giacomo and Xu, Shuqi and Mariani, Manuel S and Medo, Mat{\'u}{\v{s}}},
  journal={Proceedings of the National Academy of Sciences},
  volume={121},
  number={41},
  pages={e2410021121},
  year={2024},
  publisher={National Academy of Sciences}
}

@article{radicchi2012reverse,
  title={A reverse engineering approach to the suppression of citation biases reveals universal properties of citation distributions},
  author={Radicchi, Filippo and Castellano, Claudio},
  journal={PLoS One},
  volume={7},
  number={3},
  pages={e33833},
  year={2012},
  publisher={Public Library of Science San Francisco, USA}
}

@article{pitak2012morality,
  title={Morality, ethics, norms and research misconduct},
  author={Pitak-Arnnop, Poramate and Dhanuthai, Kittipong and Hemprich, Alexander and Pausch, Niels C},
  journal={Journal of Conservative Dentistry and Endodontics},
  volume={15},
  number={1},
  pages={92--93},
  year={2012},
  publisher={Medknow}
}

@article{bammer2008enhancing,
  title={Enhancing research collaborations: Three key management challenges},
  author={Bammer, Gabriele},
  journal={Research Policy},
  volume={37},
  number={5},
  pages={875--887},
  year={2008},
  publisher={Elsevier}
}

@article{culbert2025reference,
  title={Reference coverage analysis of OpenAlex compared to Web of Science and Scopus},
  author={Culbert, Jack H and Hobert, Anne and Jahn, Najko and Haupka, Nick and Schmidt, Marion and Donner, Paul and Mayr, Philipp},
  journal={Scientometrics},
  volume={130},
  number={4},
  pages={2475--2492},
  year={2025},
  publisher={Springer}
}

@article{radicchi2011citation,
  title={Citation networks},
  author={Radicchi, Filippo and Fortunato, Santo and Vespignani, Alessandro},
  journal={Models of science dynamics: Encounters between complexity theory and information sciences},
  pages={233--257},
  year={2011},
  publisher={Springer}
}

@article{chen2010community,
  title={Community structure of the physical review citation network},
  author={Chen, Pu and Redner, Sidney},
  journal={Journal of Informetrics},
  volume={4},
  number={3},
  pages={278--290},
  year={2010},
  publisher={Elsevier}
}

@book{zhao2015analysis,
  title={Analysis and visualization of citation networks},
  author={Zhao, Dangzhi and Strotmann, Andreas},
  year={2015},
  publisher={Morgan \& Claypool Publishers}
}

@article{braun1995scientometric,
  title={The scientometric weight of 50 nations in 27 science areas, 1989--1993. Part I. All fields combined, mathematics, engineering, chemistry and physics},
  author={Braun, Tibor and Gl{\"a}nzel, Wolfgang and Grupp, Hariolf},
  journal={Scientometrics},
  volume={33},
  pages={263--293},
  year={1995},
  publisher={Kluwer Academic Publishers}
}

@book{kuhn1997structure,
  title={The structure of scientific revolutions},
  author={Kuhn, Thomas S},
  volume={962},
  year={1997},
  publisher={University of Chicago press Chicago}
}

@article{hidalgo2009building,
  title={The building blocks of economic complexity},
  author={Hidalgo, C{\'e}sar A and Hausmann, Ricardo},
  journal={Proceedings of the national academy of sciences},
  volume={106},
  number={26},
  pages={10570--10575},
  year={2009},
  publisher={National Acad Sciences}
}

@article{sciarra2020reconciling,
  title={Reconciling contrasting views on economic complexity},
  author={Sciarra, Carla and Chiarotti, Guido and Ridolfi, Luca and Laio, Francesco},
  journal={Nature communications},
  volume={11},
  number={1},
  pages={3352},
  year={2020},
  publisher={Nature Publishing Group UK London}
}

@article{hirsch2005index,
  title={An index to quantify an individual's scientific research output},
  author={Hirsch, Jorge E},
  journal={Proceedings of the National academy of Sciences},
  volume={102},
  number={46},
  pages={16569--16572},
  year={2005},
  publisher={National Acad Sciences}
}

@book{price1963little,
  title={Little science, big science},
  author={Price, Derek J De Solla},
  year={1963},
  publisher={Columbia university press}
}

@article{bornmann2021growth,
  title={Growth rates of modern science: a latent piecewise growth curve approach to model publication numbers from established and new literature databases},
  author={Bornmann, Lutz and Haunschild, Robin and Mutz, R{\"u}diger},
  journal={Humanities and Social Sciences Communications},
  volume={8},
  number={1},
  pages={1--15},
  year={2021},
  publisher={Palgrave}
}

@article{van2004measuring,
  title={Measuring science. Capita selecta of current main issues},
  author={Van Rann, A},
  journal={Handbook of quantitative science and technology research},
  pages={19--50},
  year={2004},
  publisher={Kluwer Academic Dordrecht, the Netherlands}
}

@article{radicchi2008universality,
  title={Universality of citation distributions: Toward an objective measure of scientific impact},
  author={Radicchi, Filippo and Fortunato, Santo and Castellano, Claudio},
  journal={Proceedings of the National Academy of Sciences},
  volume={105},
  number={45},
  pages={17268--17272},
  year={2008},
  publisher={National Acad Sciences}
}

@article{bornmann2008citation,
  title={What do citation counts measure? A review of studies on citing behavior},
  author={Bornmann, Lutz and Daniel, Hans-Dieter},
  journal={Journal of documentation},
  volume={64},
  number={1},
  pages={45--80},
  year={2008},
  publisher={Emerald Group Publishing Limited}
}

@article{dworking2019emergent,
  title={The emergent integrated network structure of scientific research},
  author={Dworkin, Jordan D and Shinohara, Russell T and Bassett, Danielle S},
  journal={PloS one},
  volume={14},
  number={4},
  pages={e0216146},
  year={2019},
  publisher={Public Library of Science San Francisco, CA USA}
}

@article{ioannidis2016multiple,
  title={Multiple citation indicators and their composite across scientific disciplines},
  author={Ioannidis, John PA and Klavans, Richard and Boyack, Kevin W},
  journal={PLoS biology},
  volume={14},
  number={7},
  pages={e1002501},
  year={2016},
  publisher={Public Library of Science San Francisco, CA USA}
}

@article{bornmann2020should,
  title={Should citations be field-normalized in evaluative bibliometrics? An empirical analysis based on propensity score matching},
  author={Bornmann, Lutz and Haunschild, Robin and Mutz, R{\"u}diger},
  journal={Journal of Informetrics},
  volume={14},
  number={4},
  pages={101098},
  year={2020},
  publisher={Elsevier}
}

@article{bornmann2019normalisation,
  title={Normalisation of citation impact in economics},
  author={Bornmann, Lutz and Wohlrabe, Klaus},
  journal={Scientometrics},
  volume={120},
  number={2},
  pages={841--884},
  year={2019},
  publisher={Springer}
}

@book{moed2005citation,
  title={Citation analysis in research evaluation},
  author={Moed, Henk F},
  year={2005},
  publisher={Springer}
}

@article{ke2023network,
  title={A network-based normalized impact measure reveals successful periods of scientific discovery across disciplines},
  author={Ke, Qing and Gates, Alexander J and Barab{\'a}si, Albert-L{\'a}szl{\'o}},
  journal={Proceedings of the National Academy of Sciences},
  volume={120},
  number={48},
  pages={e2309378120},
  year={2023},
  publisher={National Acad Sciences}
}

@article{funk2017dynamic,
  title={A dynamic network measure of technological change},
  author={Funk, Russell J and Owen-Smith, Jason},
  journal={Management science},
  volume={63},
  number={3},
  pages={791--817},
  year={2017},
  publisher={INFORMS}
}

@article{park2023papers,
  title={Papers and patents are becoming less disruptive over time},
  author={Park, Michael and Leahey, Erin and Funk, Russell J},
  journal={Nature},
  volume={613},
  number={7942},
  pages={138--144},
  year={2023},
  publisher={Nature Publishing Group UK London}
}

@article{wu2019large,
  title={Large teams develop and small teams disrupt science and technology},
  author={Wu, Lingfei and Wang, Dashun and Evans, James A},
  journal={Nature},
  volume={566},
  number={7744},
  pages={378--382},
  year={2019},
  publisher={Nature Publishing Group UK London}
}

@article{kuhn1970nature,
  title={The nature of scientific revolutions},
  author={Kuhn, Thomas},
  journal={Chicago: University of Chicago},
  volume={197},
  number={0},
  year={1970}
}

@article{leydesdorff2016citations,
  title={Citations: Indicators of quality? The impact fallacy},
  author={Leydesdorff, Loet and Bornmann, Lutz and Comins, Jordan A and Milojevi{\'c}, Sta{\v{s}}a},
  journal={Frontiers in Research metrics and Analytics},
  volume={1},
  pages={1},
  year={2016},
  publisher={Frontiers Media SA}
}

@article{forthmann2024summing,
  title={Why summing up bibliometric indicators does not justify a composite indicator},
  author={Forthmann, Boris and Doebler, Philipp and Mutz, R{\"u}diger},
  journal={Scientometrics},
  volume={129},
  number={12},
  pages={7475--7499},
  year={2024},
  publisher={Springer}
}

@article{kadanoff1966scaling,
  title={Scaling laws for Ising models near T c},
  author={Kadanoff, Leo P},
  journal={Physics Physique Fizika},
  volume={2},
  number={6},
  pages={263},
  year={1966},
  publisher={APS}
}

@article{wilson1971renormalizationa,
  title={Renormalization group and critical phenomena. I. Renormalization group and the Kadanoff scaling picture},
  author={Wilson, Kenneth G},
  journal={Physical review B},
  volume={4},
  number={9},
  pages={3174},
  year={1971},
  publisher={APS}
}

@article{wilson1971renormalizationb,
  title={Renormalization group and critical phenomena. II. Phase-space cell analysis of critical behavior},
  author={Wilson, Kenneth G},
  journal={Physical Review B},
  volume={4},
  number={9},
  pages={3184},
  year={1971},
  publisher={APS}
}

@article{higgs1964broken,
  title={Broken symmetries and the masses of gauge bosons},
  author={Higgs, Peter W},
  journal={Physical review letters},
  volume={13},
  number={16},
  pages={508},
  year={1964},
  publisher={APS}
}

@article{baibich1988giant,
  title={Giant magnetoresistance of (001) Fe/(001) Cr magnetic superlattices},
  author={Baibich, Mario Norberto and Broto, Jean Marc and Fert, Albert and Van Dau, F Nguyen and Petroff, Fr{\'e}d{\'e}ric and Etienne, P and Creuzet, G and Friederich, A and Chazelas, J},
  journal={Physical review letters},
  volume={61},
  number={21},
  pages={2472},
  year={1988},
  publisher={APS}
}

@article{jurgen2012francis,
  title={Francis Bacon},
  author={J{\"u}rgen, Klein},
  journal={Edited by Edward N. Zalta. The Stanford Encyclopedia of Philosophy, Winter},
  year={2012}
}

@article{shibayama2021measuring,
  title={Measuring novelty in science with word embedding},
  author={Shibayama, Sotaro and Yin, Deyun and Matsumoto, Kuniko},
  journal={PloS one},
  volume={16},
  number={7},
  pages={e0254034},
  year={2021},
  publisher={Public Library of Science San Francisco, CA USA}
}

@article{leahey2023types,
  title={What types of novelty are most disruptive?},
  author={Leahey, Erin and Lee, Jina and Funk, Russell J},
  journal={American Sociological Review},
  volume={88},
  number={3},
  pages={562--597},
  year={2023},
  publisher={SAGE Publications Sage CA: Los Angeles, CA}
}

@article{redner2005citation,
  title={Citation statistics from 110 years of physical review},
  author={Redner, Sidney},
  journal={Physics today},
  volume={58},
  number={6},
  pages={49--54},
  year={2005},
  publisher={AIP Publishing}
}

\setcounter{figure}{0}
\setcounter{table}{0}
\renewcommand{\thefigure}{S\arabic{figure}}
\renewcommand{\thetable}{S\arabic{table}}

\clearpage
\section*{Supplementary Information (SI)}

This supplementary information provides additional details on the results, document data collection, methodological refinements, and additional results supporting the findings reported in the main article.

\subsection*{Network-based normalized citation (NBNC) score}
From the citation network constructed from the dataset, in order to identify breakthrough publications, we first compute the NBNC score \cite{ke2023network}. In the NBNC approach, the raw citation count of a focal paper is compared with the average citations of its co-cited publications for each year $t$ till the time-horizon $\tau$, after its publication. For instance, let $f$ be the focal paper. Consider $\Gamma_f(t) = |\mathrm{Citations}_f(t)|$ as the total number of citations received by $f$ at citation age $t$, where $\mathrm{Citations}_f(t)$ is the set of all the papers citing $f$. Also, $\mathrm{CoCited}_f(t)$ is the set of papers cocited with $f$ at $t$. The NBNC score for paper $f$ at an age horizon $\tau$ may be computed as:
\begin{equation*}
\mathrm{NBNC}_f(\tau) = \sum_{t=0}^{\tau} \frac{\Gamma_f(t)}{\left\langle \Gamma_j(t) \right\rangle_{j \in \mathrm{CoCited}_f(t)}}.
\end{equation*}
Here, the denominator is the average $\Gamma_j(t)$ of the co-cited papers $j \in \mathrm{CoCited}_f(t)$ at $j$'s citation age $t$ and normalizes $\Gamma_f(t)$. This is summed over for $\tau = 10$ years to construct the score. We recognise $f$ as a \emph{breakthrough} if it ranks in the top 5\% of NBNC scores among all the papers from the same publication year $y$. Using the labels that OpenAlex provided, we then categorise these breakthrough papers into different subfields. $\mathrm{BT}_{s}(y)$ represents the set of all breakthrough papers from a subfield $s$ for a publication year $y$. The number of these papers in each $s$ for each $y$ is represented by $N_s^{\mathrm{BT}}(y) = |BT_s(y)|$.

\subsection*{The Disruptive Index (CD)}
These breakthrough papers are further divided into two based on their consolidating or disruptive nature using \emph{Disruptive index} \cite{funk2017dynamic,wu2019large}. This index captures the local structural change induced by the introduction of a node into the citation network. For its computation, we start by partitioning the citations of a paper $f$ at the horizon $\tau$ ($\mathrm{Citations}_f(\tau)$) into two disjoint subsets. Citation of $f$ that does not cite any of $f$'s references; and citations of $f$ that cite at least one of $f$'s references. If we define $\mathrm{References}_f$ as the set of papers cited by a paper $f$, their cardinality are represented by:
\begin{align*}
\Gamma^{\mathrm{exc}}_f(\tau) = \left| \left\{
    j \in \mathrm{Citations}_f(\tau) :\ 
    \mathrm{References}_j \cap \mathrm{References}_f = \emptyset
\right\} \right|, \\
\Gamma^{\mathrm{inc}}_f(\tau) = \left| \left\{
    j \in \mathrm{Citations}_f(\tau) :\ 
    \mathrm{References}_j \cap \mathrm{References}_f \neq \emptyset
\right\} \right|.
\end{align*}
These represent the number of \emph{exclusive} and \emph{inclusive} citations to $f$ respectively till the citation age $\tau$, and satisfy the identity:
$
\Gamma_f(\tau) = \Gamma_f^{\mathrm{exc}}(\tau) + \Gamma_f^{\mathrm{inc}}(\tau).
$
We also take the set of all papers that cite at least one reference of $f$ but not $f$ itself, within $\tau$ years of its publication, and define it cardinality as:
\begin{align*}
\Gamma^{\mathrm{ref}}_f(\tau) = \left| \left\{
    j \notin \mathrm{Citations}_f(\tau) :\ 
    \mathrm{References}_j \cap \mathrm{References}_f \neq \emptyset
\right\} \right|,
\end{align*}
Using them, the CD index of $f$ at $\tau$ is computed as:
\begin{equation*}
\mathrm{CD}_f(\tau) = \frac{\Gamma_f^{\mathrm{exc}}(\tau) - \Gamma_f^{\mathrm{inc}}(\tau)}{\Gamma_f(\tau) + \Gamma_f^{\mathrm{ref}}(\tau)}.
\end{equation*}
A positive value of $\mathrm{CD}_f(\tau)$ indicates that $f$ is cited independently of its references (\emph{disruptive}), while a negative value indicates that it is cited alongside its references (\emph{consolidating}).

After applying this classification to every breakthrough paper, we divide $\mathrm{BT}_s(y)$ into consolidating breakthrough ($\mathrm{CN}_s(y) = \left\{
    j \in \mathrm{BT}_s(y) :\ \mathrm{CD}_j(\tau) \leq 0 \right\} $)
and disruptive breakthroughs ($\mathrm{DI}_s(y) = \left\{
    j \in \mathrm{BT}_s(y) :\ \mathrm{CD}_j(\tau) > 0 \right\}$);
 with their respective counts $N_s^{\mathrm{CN}}(y) = |\mathrm{CN}_s(y)|$ and $N_s^{\mathrm{DI}}(y) = |\mathrm{DI}_s(y)|$. Here, $\mathrm{BT}_s(y) = \mathrm{CN}_s(y) \cup \mathrm{DI}_s(y)$, $\mathrm{CN}_s(y) \cap \mathrm{DI}_s(y) = \emptyset$, and $
N_s^{\mathrm{BT}}(y) = N_s^{\mathrm{CN}}(y) + N_s^{\mathrm{DI}}(y),
$. The evolution of $N_s^{\mathrm{CN}}(y)$ and $N_s^{\mathrm{DI}}(y)$ of every subfield $s \in \ \mathcal{S}$ is plotted in Figure~\ref{fig:timeseries_BT_DI}. 

\subsection*{Dynamic time warping (DTW) and Gaussian kernel}

Now we will be comparing the dynamics of different subfields through this space of consolidating and disruptive breakthroughs. Instead of directly using the time series of absolute counts of these breakthroughs,
$
\left\{ \left( N^{CN}_s(y), N^{DI}_s(y) \right) \mid y \in \mathcal{Y} \right\},
$
We first normalise these counts by the total number of papers published in the respective subfields for each year. For each subfield \(s \in S\), where \(S \subset \mathcal{S}\) denotes the set of 61 selected subfields (see Results), we define the 2-dimensional time series,
\[
\mathcal{T}_s = \left\{ \left( \widetilde{N}^{CN}_s(y), \widetilde{N}^{DI}_s(y) \right) \mid y \in \mathcal{Y} \right\}.
\]

Here,
$
\tilde{N}^{\mathrm{CN}}_s(y) = \frac{N^{\mathrm{CN}}_s(y)}{N_s(y)}$ and 
$
\tilde{N}^{\mathrm{DI}}_s(y) = \frac{N^{\mathrm{DI}}_s(y)}{N_s(y)},
$
represent the fraction of consolidating and disruptive breakthroughs in subfield \(s\) during year \(y\), and \(\mathcal{Y}\) denotes the set of publication years considered. To compare these temporal profiles of two subfields and calculate a distance, we use Dynamic Time Warping (DTW)\cite{keogh2005exact} and measure the alignment cost between their trajectories.

For each pair of subfields \(s_i, s_j \in S\), we start by computing a cost matrix between all the corresponding time point pairs \(y_k, y_l \in \mathcal{Y}\). 
$$
D_{s_i, s_j}(y_k, y_l) =
\left( \widetilde{N}^{CN}_{s_i}(y_k) - \widetilde{N}^{CN}_{s_j}(y_l) \right)^2
+
\left( \widetilde{N}^{DI}_{s_i}(y_k) - \widetilde{N}^{DI}_{s_j}(y_l) \right)^2
$$
Using this squared Euclidean distance matrix as the cost matrix, we find the optimal warping path \(W\): a sequence of index pairs \((y_k, y_l)\) satisfying boundary and monotonicity conditions that define a valid alignment, which minimized the cost function $\sum_{(y_k, y_l) \in W} D_{s_i, s_j}(y_k, y_l)$. The DTW distance between \(s_i\) and \(s_j\) is given by the minimum value of the cost function from $W$.
\[
\text{DTW}_{s_i, s_j} = \min_{W} \sum_{(y_k, y_l) \in W} D_{s_i, s_j}(y_k, y_l).
\]
To convert the resulting distance matrix into a similarity matrix, we apply a Gaussian kernel transformation:
\[
\mathcal{K}_{s_i, s_j} = \exp\left( -\frac{ \text{DTW}_{s_i, s_j}^2 }{ 2\sigma^2 } \right),
\]
where \(\sigma\) is a scaling parameter set to the median of all pairwise DTW distances. The resulting matrix \(\mathcal{K}\) defines a fully connected, weighted graph over subfields in \(S\), where edge weights reflect the temporal similarity in their breakthrough trajectories. We apply the Leiden algorithm to this graph to identify clusters of subfields with similar historical patterns in the evolution of consolidating and disruptive scientific activity. The Leiden Algorithm is an advanced community detection algorithm that improves upon the Louvain algorithm, ensuring well-connected internal communities.  It consists of three main phases: Local moving of nodes, refinement of the partition, and aggregation of the network \cite{traag2019louvain}.

\subsection*{GENEPY}\label{rca_genepy}
We start by constructing $\textbf{P}^\alpha(y)\in \mathbb{N}^{C \times S}$ for all years $y \in \mathcal{Y}$ and breakthrough type $\alpha \in \{\mathrm{CN}, \mathrm{DI}\}$ by counting number of $\alpha$ type breakthroughs in each subfield $s \in \mathcal{S}$ attributed to each country $c \in \mathcal{C}$. For the purpose of this study, the countries are assigned by giving full credit to every country affiliated with the paper. We then aggregate these yearly matrices to $\textbf{P}^\alpha(d) \in \mathbb{N}^{C \times S}$ where $d \in D$ are different decades. Their elements are related by $P^\alpha_{cs}(d) = \sum_{y \in d} P^\alpha_{cs}(y)$.

To account for the size effects among different subfields we normalize $\textbf{P}^\alpha(d)$ to obtain $\tilde{\textbf{P}}^\alpha(d)$ as $\tilde{P}^\alpha_{cs}(d) = \frac{P^\alpha_{cs}(d)}{N_s(d)}$, where $N_s(d) =\sum_{y \in d} N_s(y)$ denotes the total number of breakthroughs in subfield $s$ in decade $d$.

To measure the significance of the contribution of a country to a subfield, we compute the revealed comparative advantage (RCA)\cite{balassa1965trade} for each country-subfield pair, constructing $\mathrm{\textbf{RCA}}^\alpha(d)$:
$
\mathrm{RCA}^\alpha_{cs}(d) = \frac{\tilde{P}^\alpha_{cs}(d)}{\sum_{s'} \tilde{P}^\alpha_{cs'}(d)} \Big/ \frac{\sum_{c'} \tilde{P}^\alpha_{c's}(d)}{\sum_{c',s'} \tilde{P}^\alpha_{c's'}(d)}.
$
By putting a threshold we can then define the binary adjacency matrix $\textbf{A}^\alpha(d) \in \{0, 1\}^{C \times S}$, where $A^\alpha_{cs}(d) = 1$ if $\mathrm{RCA}^\alpha_{cs}(d) > 1$, and zero otherwise. 
Giving us the bipartite network representation. This allows us to apply economic complexity formalism to obtain two sets of scores for the countries and subfields, respectively.

We proceed by defining a rationalized matrix $\widetilde{\textbf{A}}^\alpha(d) \in \mathbb{R}^{C \times S}$ with entries

$$
\widetilde{A}^\alpha_{cs}(d) = \frac{A^\alpha_{cs}(d)}{k^\alpha_c(d) \cdot k'^\alpha_s(d)},
$$

\noindent where $k^\alpha_c(d) = \sum_s A^\alpha_{cs}(d)$ is the number of subfields in which country $c$ is active, and $k'^\alpha_s(d) = \sum_c A^\alpha_{cs}(d)/k^\alpha_c(d)$ is a weighted sum over countries contributing to subfield $s$, where each country is weighted inversely by the number of subfields it is active in. This particular normalisation structure is chosen to capture the iterative relationship: "The more complex countries are the ones that perform well in more complex subfields, and more complex subfields are the ones that only the more complex countries can perform". 
The mathematical formalism originally given by \textbf{Tachella et. al}  was linearised by \textbf{GENEPY group} by taking a first-order approximation. This allows us to compute the two sets of scores from the leading eigenvectors of two similarity matrices
$ \textbf{U}^\alpha(d) = \widetilde{\textbf{A}}^\alpha(d)\, \widetilde{\textbf{A}}^\alpha(d)^\top$ and $ \quad \textbf{V}^\alpha(d) = \widetilde{\textbf{A}}^\alpha(d)^\top\, \widetilde{\textbf{A}}^\alpha(d),$ between countries and subfields respectively. After we perform eigen decomposition of both matrices and retain the top two eigenvectors, each country $c$ is assigned a score:

$$
H^\alpha_c(d) = \sqrt{ \left( u^{(1)}_c \right)^2 + \left( u^{(2)}_c \right)^2 }, \quad
G^\alpha_s(d) = \sqrt{ \left( v^{(1)}_s \right)^2 + \left( v^{(2)}_s \right)^2 },
$$

\noindent where $\textbf{u}^{(1)}$ and $\textbf{u}^{(2)}$ are the leading eigenvectors of $\textbf{U}^\alpha(d)$, and similarly $\textbf{v}^{(1)}$, $\textbf{v}^{(2)}$ are of $\textbf{V}^\alpha(d)$. These scores are then used for ranking the subfields and countries, and the evolution of these ranks over the 7 decades is shown in fig. \ref{fig:Allcountry_ranks}

\vspace{0.4cm}

\begin{definition}\label{hindex}
\textbf{The h-index} 

J.E. Hirch introduced the $h$-index \cite{hirsch2005index}, a widely used metric to quantify scientific performance. The index is defined as follows: an actor(author/country/institution) has an $h$-index if $h$ papers have been cited at least $h$ times. For example, if a country has $10$ publications and each publication has been cited at least 10 times, then the country has an $h$-index of 10. 
\end{definition}

\begin{figure}[htbp]
\centering
\includegraphics[width=0.65\linewidth]{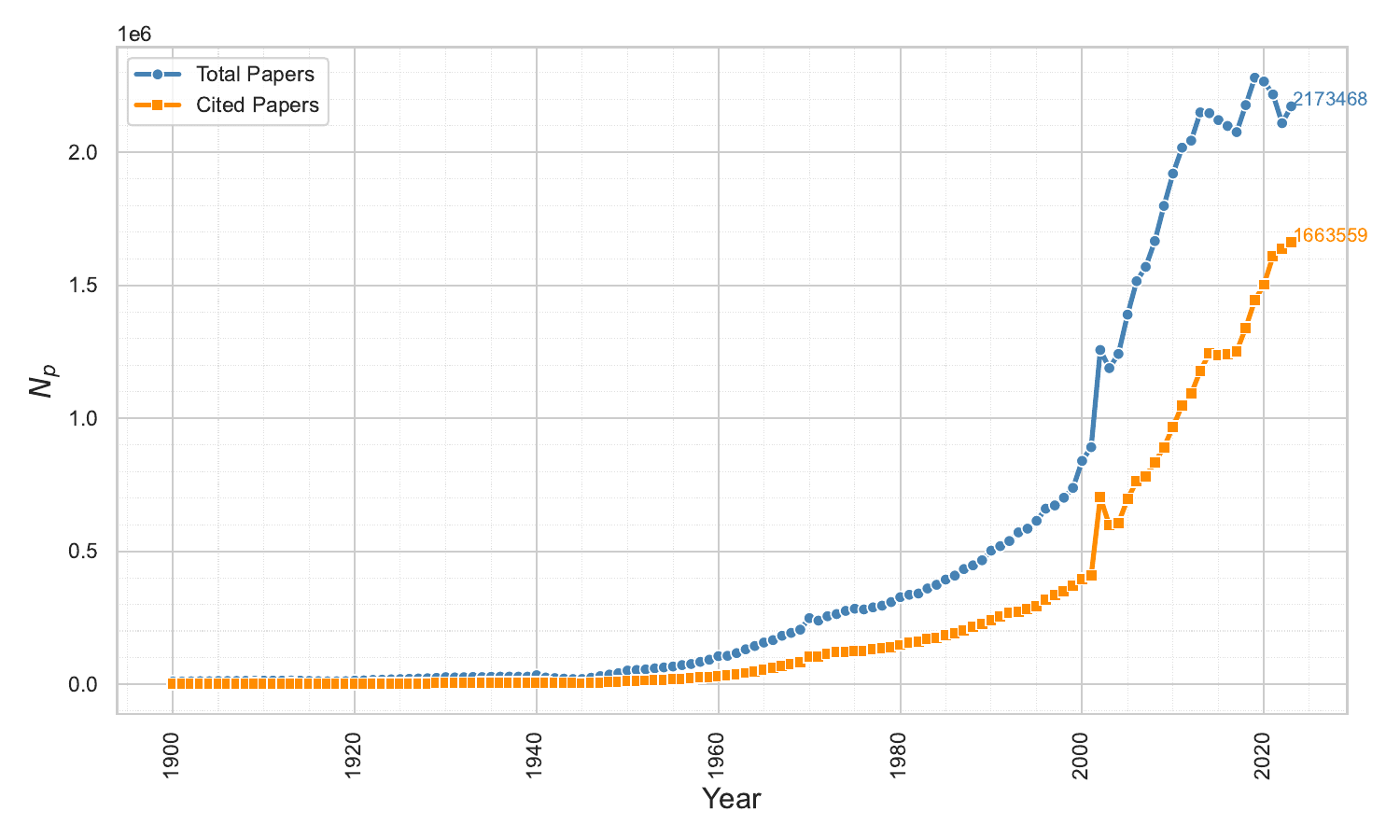}
\caption{Yearly growth of papers represented in blue color, while the yearly growth of cited papers is shown in orange color.}
\label{fig:PaperGDynamics}
\end{figure}

We identified 61 subfields, out of 89, based on their sustained growth in publication volume and proportion of breakthroughs, to construct the complexity ranking of countries and subfields. Abbreviations used in the rankings are listed with the corresponding subfield names.

\begin{table}[htbp]
\hspace*{-1cm}
    \centering
    \begin{tabular}{|c|c|c|c|}

    \hline
    Sr. No. & Subfields(Abbreviation)&Sr. No. & Subfields(Abbreviation)\\ 
    \hline
        1&Aerospace Engineering (ASE) & 32&Geophysics(GP)  \\
2&Algebra and Number Theory (ANT) &  33&Global and Planetary Change (GPC)\\
3&Analytical Chemistry (AC) & 34&Health, Toxicology and Mutagenesis(HTM) \\
4&Applied Mathematics (AM) &35&Industrial and Manufacturing Engineering(IME) \\
5&Artificial Intelligence (AI) &36&Inorganic Chemistry (IC)  \\
6&Atmospheric Science (AS)&  37&Information Systems (IS)\\
7&Astronomy and Astrophysics (AA), &  38&Materials Chemistry (MC)\\
8&Atomic and Molecular Physics and Optics (AMPO)&39&Mathematical Physics (MP)\\
9&Automotive Engineering (AME) &40&Management, Monitoring, Policy and Law (MMPL)\\
10&Biomaterials (BM) &41&Mechanical Engineering (ME)\\\
11&Biomedical Engineering (BE) &  42&Mechanics of Materials (MM)\\
12&Building and Construction (BC) & 43&Nature and Landscape Conservation (NLC)\\
13&Civil and Structural Engineering (CSE-1) &44&Nuclear and High Energy Physics (NHEP) \\
14&Computational Mechanics (CM)&45&Numerical Analysis (NA)\\

15&Computational Theory and Mathematics (CTM) &46&Oceanography (OG)\\
16&Computer Networks and Communications (CNC) &  47&Ocean Engineering (OE) \\
17&Computer Science Applications (CSA) & 48&Organic Chemistry (OC)  \\
18&Computer Vision and Pattern Recognition (CVPR) &49&Pollution (PL)  \\
19&Condensed Matter Physics (CMP)&50&Paleontology (PT)\\
20&Control and Systems Engineering (CSE-2)&51&Polymers and Plastics(PP)\\
21&Discrete Mathematics and Combinatorics (DMC)&52&Physical and Theoretical Chemistry (PTC)\\
22&Electrical and Electronic Engineering (EEE)&  53&Radiation (RD)\\
 23&Ecology(EL) &54&Renewable Energy, Sustainability and the Environment (RESE)\\
24&Earth-Surface Processes (ESP)&  55&Safety, Risk, Reliability and Quality (SRRQ)\\
25&Ecological Modeling (EM) &    56&Statistics and Probability (SP)\\
26&Electronic, Optical and Magnetic Materials (EOMM)& 57&Spectroscopy (SS)   \\
27&Environmental Chemistry (EC) &58&Signal Processing (SGP)\\
28&Environmental Engineering (EE)&59&Statistical and Nonlinear Physics (SNP) \\
29&Fluid Flow and Transfer Processes(FFTP)&60&Surfaces, Coatings and Films (SCF)  \\
30&Geochemistry and Petrology (GCP)&61&Water Science and Technology(WST) \\  
31&Geometry and Topology (GT)& &\\

\hline
\end{tabular}
\caption{Selected subfields based on their growth pattern and breakthroughs evolution for the country-subfield performance matrix.}
\label{tab:selected_subfields}
\end{table}

\begin{table}[htbp]
    \centering
    \begin{tabular}{llr}
\toprule
 & Name & Cluster \\
\midrule
1 & Discrete Mathematics and Combinatorics & 1 \\
2 & Environmental Engineering & 1 \\
3 & Applied Mathematics & 1 \\
4 & Paleontology & 1 \\
5 & Algebra and Number Theory & 1 \\
6 & Mathematical Physics & 1 \\
7 & Statistics and Probability & 1 \\
8 & Astronomy and Astrophysics & 1 \\
9 & Geometry and Topology & 1 \\
10 & Biomaterials & 2 \\
11 & Pollution & 2 \\
12 & Nature and Landscape Conservation & 2 \\
13 & Health, Toxicology and Mutagenesis & 2 \\
14 & Global and Planetary Change & 2 \\
15 & Environmental Chemistry & 2 \\
16 & Analytical Chemistry & 2 \\
17 & Earth-Surface Processes & 2 \\
18 & Statistical and Nonlinear Physics & 3 \\
19 & Oceanography & 3 \\
20 & Ecology & 3 \\
21 & Biomedical Engineering & 3 \\
22 & Atmospheric Science & 3 \\
23 & Geophysics & 3 \\
24 & Materials Chemistry & 4 \\
25 & Nuclear and High Energy Physics & 4 \\
26 & Organic Chemistry & 4 \\
27 & Spectroscopy & 4 \\
28 & Physical and Theoretical Chemistry & 5 \\
29 & Electronic, Optical and Magnetic Materials & 5 \\
30 & Atomic and Molecular Physics, and Optics & 5 \\
31 & Civil and Structural Engineering & 6 \\
32 & Mechanical Engineering & 6 \\
33 & Ocean Engineering & 6 \\
34 & Geochemistry and Petrology & 7 \\
35 & Mechanics of Materials & 7 \\
36 & Computational Theory and Mathematics & 8 \\
37 & Computational Mechanics & 8 \\
38 & Control and Systems Engineering & 9 \\
39 & Artificial Intelligence & 9 \\
40 & Aerospace Engineering & 10 \\
41 & Building and Construction & 10 \\
42 & Automotive Engineering & 11 \\
43 & Information Systems & 11 \\
44 & Water Science and Technology & 12 \\
45 & Management, Monitoring, Policy and Law & 12 \\
\bottomrule
\end{tabular}
\caption{The selected subfields are clustered using the Leiden algorithm with clusters of size $ \geq 2$}
\label{tab:clustered_fields}
\end{table}

\begin{figure}[htbp]
\centering
\includegraphics[width=1\linewidth]{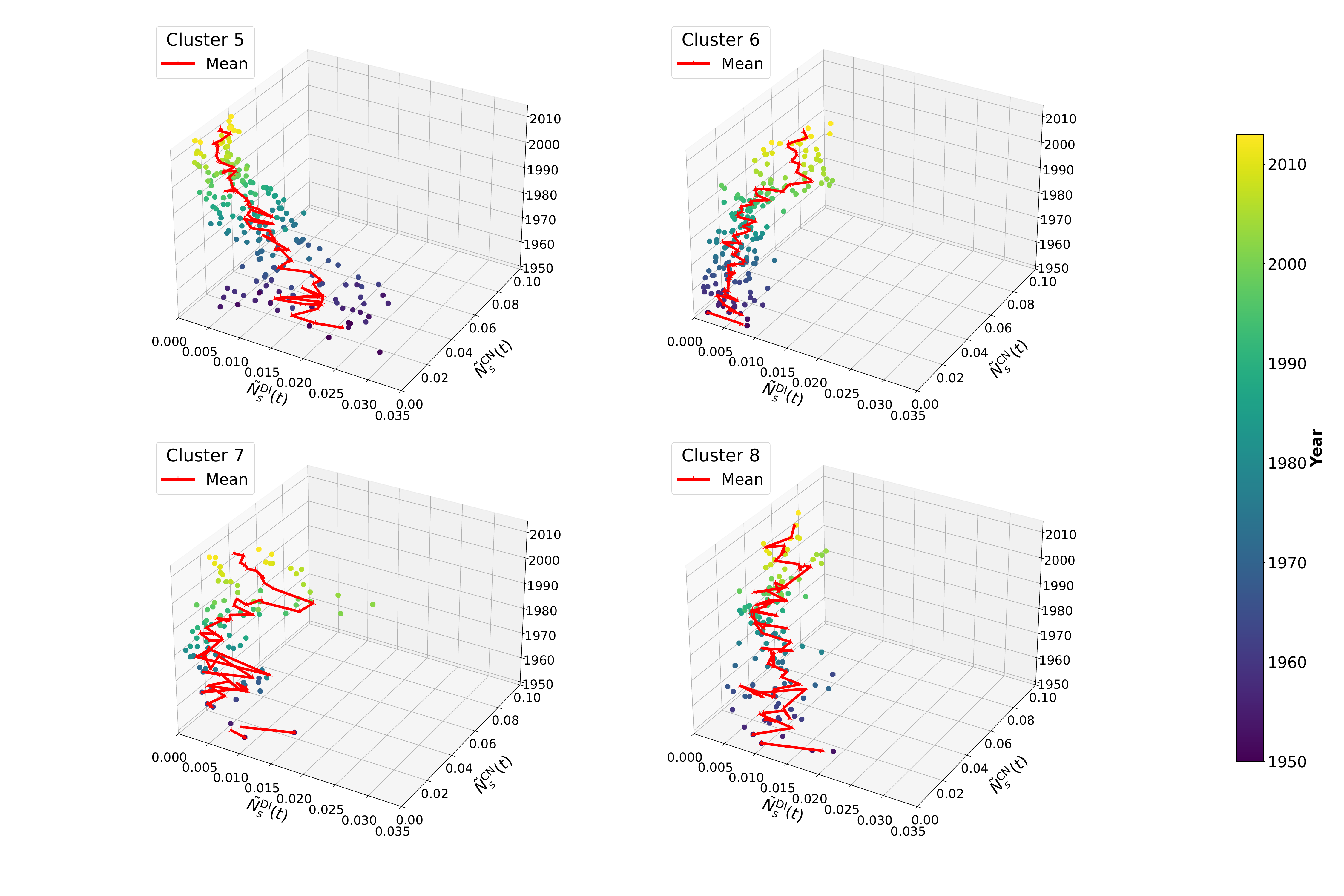}
\includegraphics[width=1\linewidth]{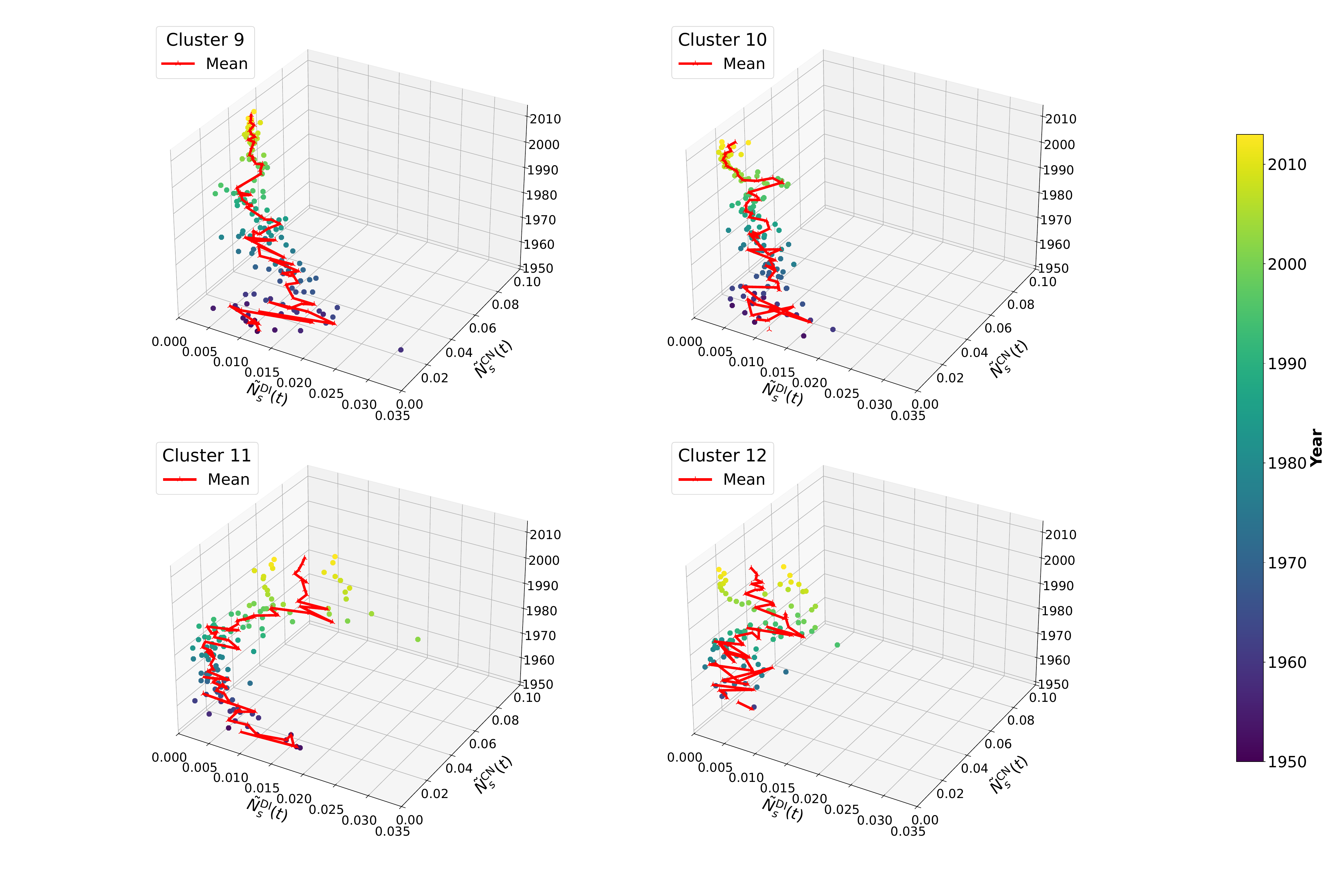}
\caption{Average growth trajectories of the eight smaller clusters, shown alongside their mean cluster trends.}
\label{fig:ClusterDynamics}
\end{figure}

\begin{figure*}[htbp]
\centering
\includegraphics[width=0.48\linewidth]{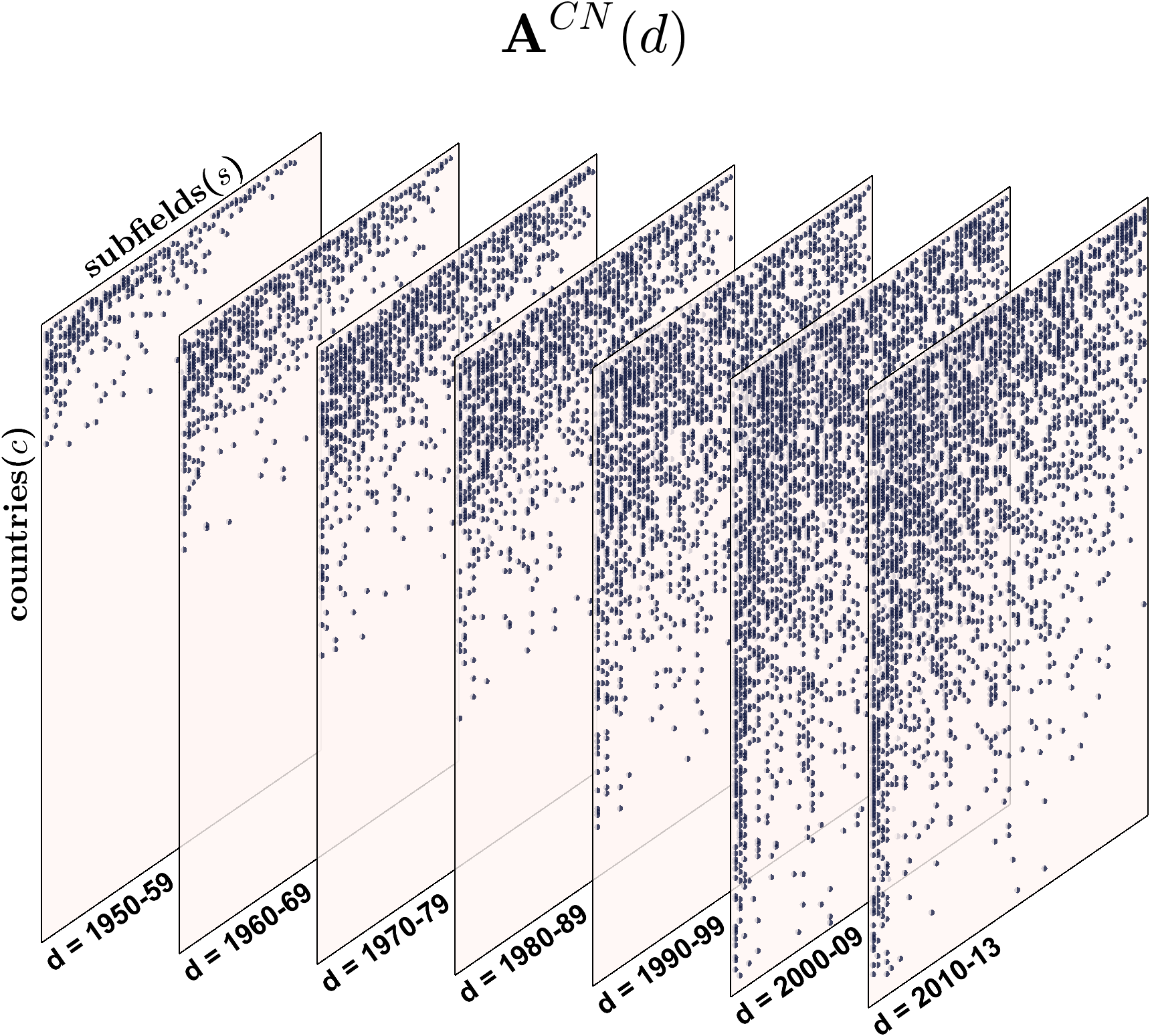}
\caption{The country-subfield RCA-filtered adjacency matrix, $\mathbf{A}^{(\text{CN})}$, for consolidating breakthroughs, shown for the  $1^{th}$, $2^{th}$, $3^{th}$, $5^{th}$, $6^{th}$, and $7^{th}$ decades (d). Countries (rows) and subfields (columns) are ordered in decreasing sequence based on the row sums (countries) and column sums (subfields), highlighting subfields in which countries gain a comparative advantage.}
\label{fig:MatrixplotCN}
\end{figure*}

\begin{figure*}[htbp]
\centering
\includegraphics[width=0.48\linewidth]{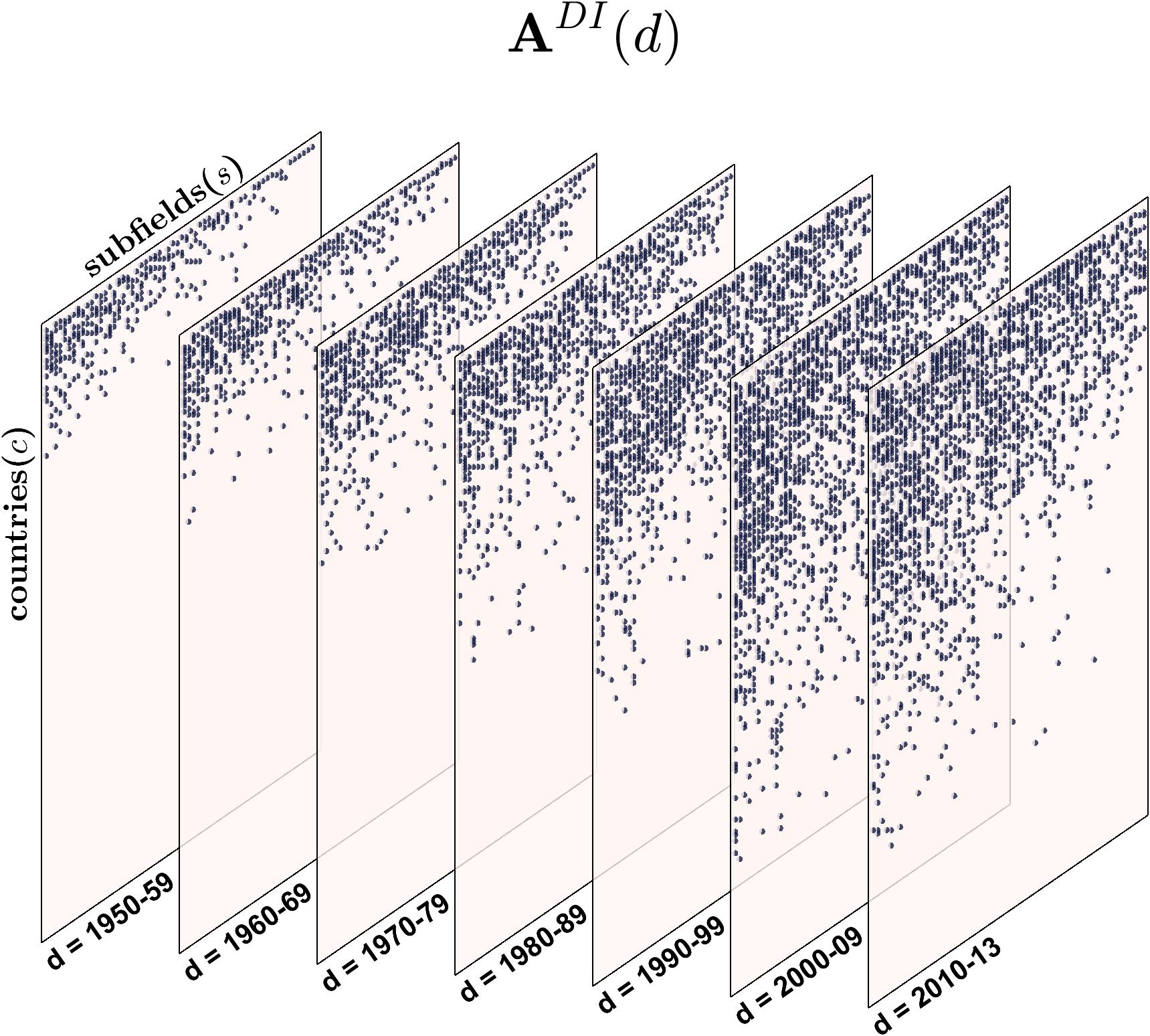}
\caption{The country-subfield adjacency matrices, $\mathbf{C}^{(\text{DI})}$ and the adjacency RCA-filtered matrix, $\mathbf{A}^{(\text{DI})}$, for disruptive breakthroughs, shown for the  $1^{th}$, $2^{th}$, $3^{th}$, $5^{th}$, $6^{th}$, and $7^{th}$ decades (d). Countries (rows) and subfields (columns) are ordered in decreasing sequence based on the row sums (countries) and column sums (subfields), highlighting subfields in which countries gain a comparative advantage.}
\label{fig:MatrixplotDI}
\end{figure*}

\begin{figure}[htbp]
  \centering
  \begin{subfigure}[t]{1\textwidth}
    \centering
    {\includegraphics[width=0.45\linewidth]{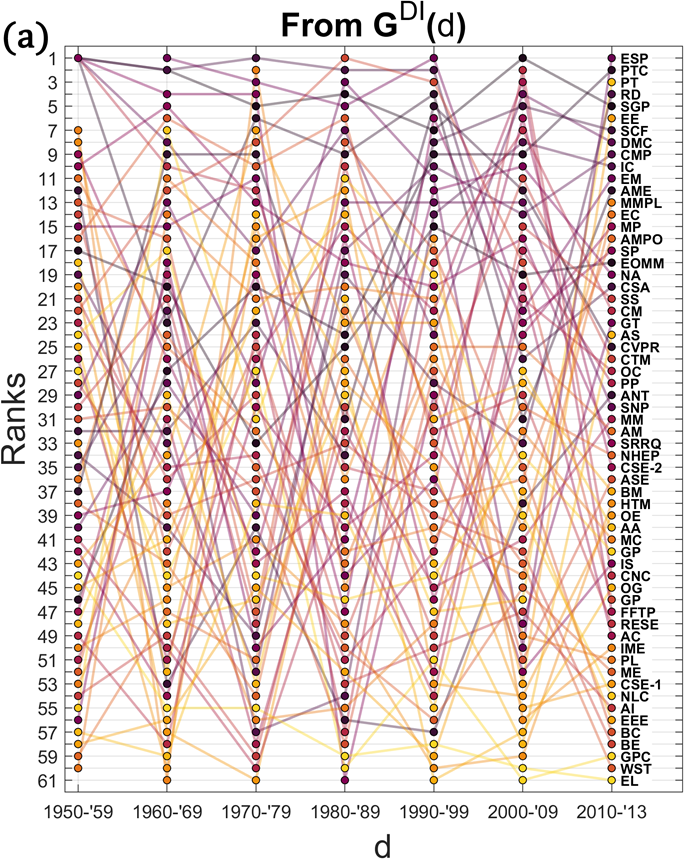}}
    \caption*{} 
    \label{fig:FieldRankingDI_All}
  \end{subfigure}
  \vspace{0.3em} 
  \begin{subfigure}[t]{1\textwidth}
    \centering
    {\includegraphics[width=0.45\linewidth]{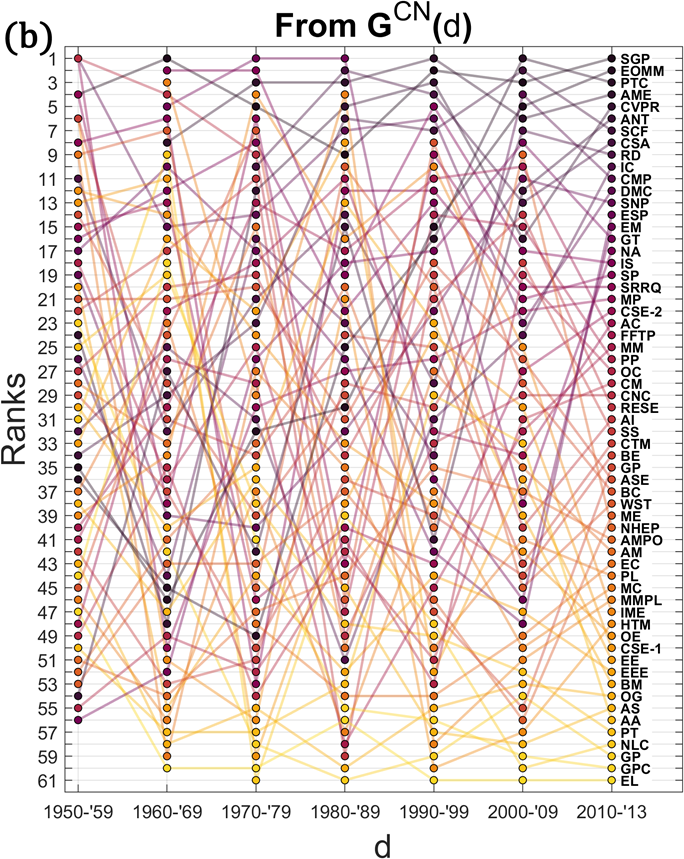}}
    \caption*{}
    \label{fig:FieldRankingCN_All}
  \end{subfigure}
  \caption{The decadal ranking of competitive subfields in disruptive and consolidating breakthroughs spans all seven decades, with the $7^{\text{th}}$ decade defined as the period from 2009 to 2013. The upper panel (a) depicts the rankings of all subfields related to disruptive breakthroughs. The lower panel (b) presents the rankings of subfields focused on consolidating breakthroughs. In the earlier decades, multiple subfields had overlapping rankings, as shown in the figure.}
  \label{fig:Decadal_Subfield_Ranking}
\end{figure}

\begin{figure}[htbp]
\centering
\includegraphics[width=1\linewidth]{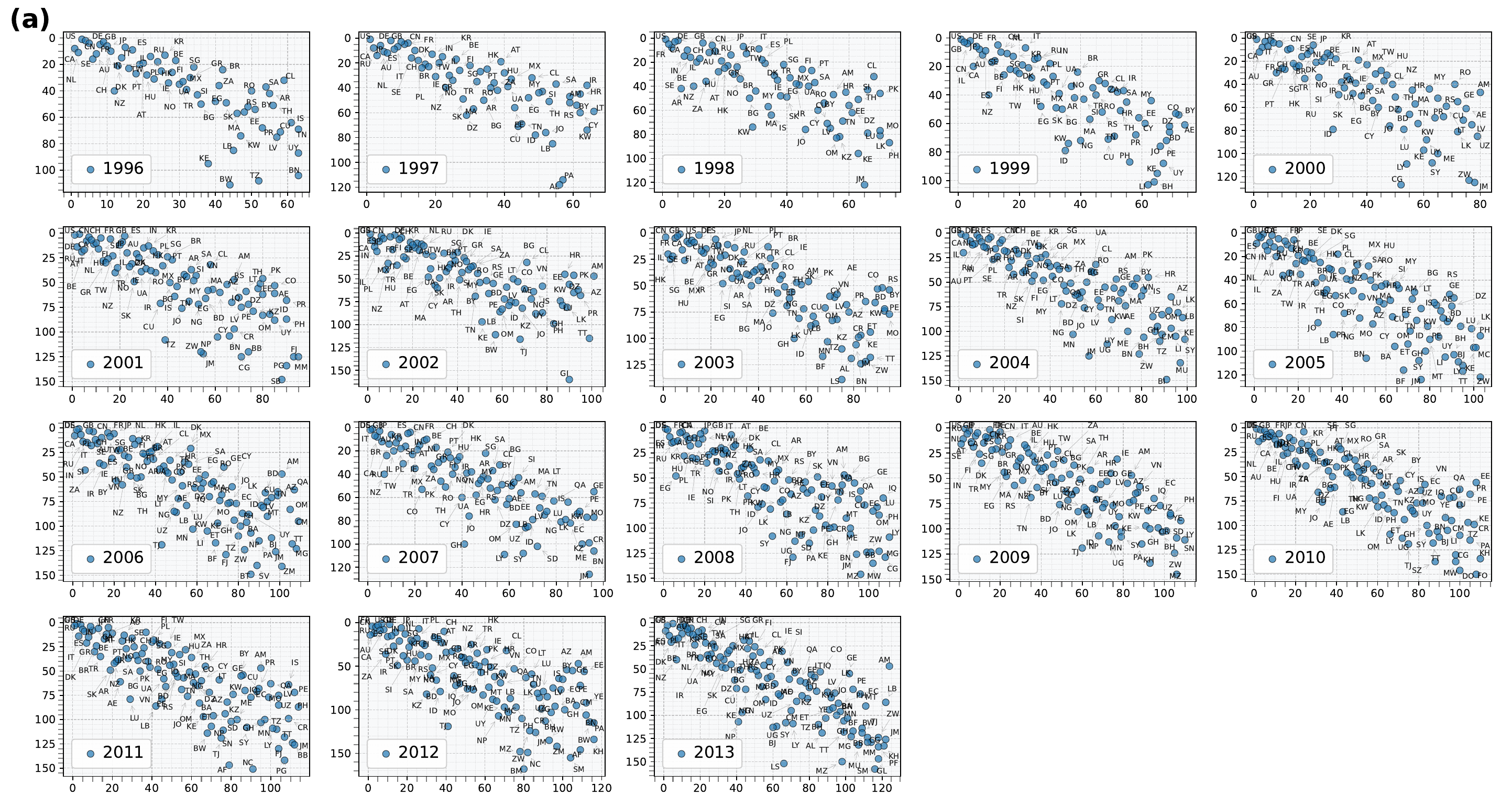}\\
\includegraphics[width=1\linewidth]{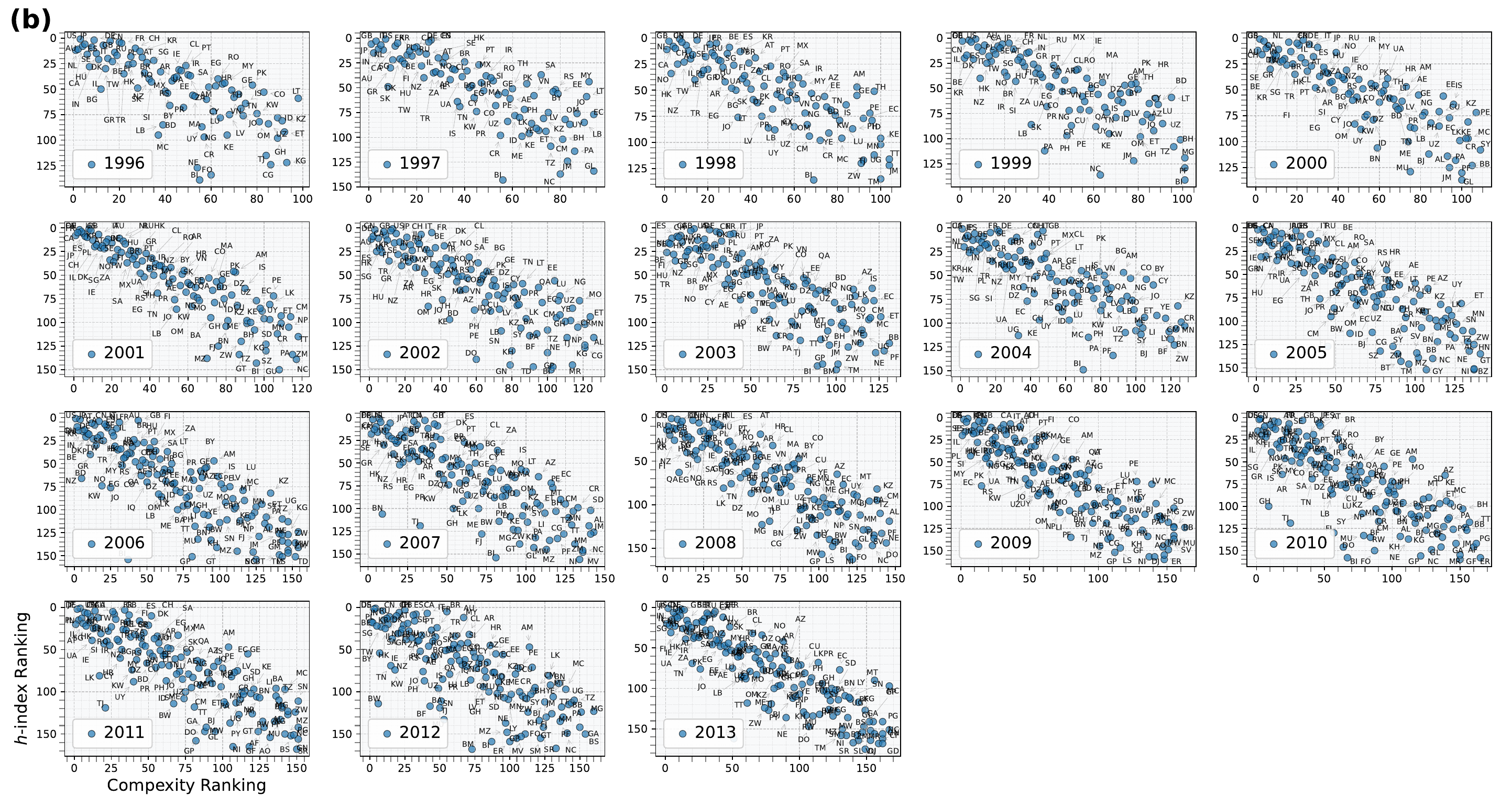}
\caption{The correlation scatter plot between ranking of the countries based on their complexity scores, {\bf (a)} for disruptive breakthroughs, while {\bf (b)} for the consolidating breakthroughs, and ranking based on the $h$-index (Scimago) from $1996$ to $2013$.
}
\label{fig:ScimagoRanking}
\end{figure}

\end{document}